\documentclass[iop, revtex4]{emulateapj}
\usepackage{multirow}
\usepackage{color}
\usepackage[normalem]{ulem}
\usepackage{amssymb}
\usepackage{lscape}

\begin{document}

\title{MIRIS P\MakeLowercase{a}$\alpha$ Galactic Plane Survey. I. Comparison with IPHAS H$\alpha$ in $\ell = 96\arcdeg$--$116\arcdeg$}

\author{Il-Joong Kim\altaffilmark{1}}
\author{Jeonghyun Pyo\altaffilmark{1}}
\author{Woong-Seob Jeong\altaffilmark{1,2}}
\author{Kwang-Il Seon\altaffilmark{1,2}}
\author{Takao Nakagawa\altaffilmark{3}}
\author{Min Gyu Kim\altaffilmark{4}}
\author{Won-Kee Park\altaffilmark{1}}
\author{Dae-Hee Lee\altaffilmark{1,2}}
\author{Dukhang Lee\altaffilmark{5}}
\author{Bongkon Moon\altaffilmark{1}}
\author{Sung-Joon Park\altaffilmark{1}}
\author{Youngsik Park\altaffilmark{1}}
\author{Toshio Matsumoto\altaffilmark{1,3}}
\author{Wonyong Han\altaffilmark{1,2}}

\altaffiltext{1}{Korea Astronomy and Space Science Institute, Daejeon, 34055, Republic of Korea; ijkim@kasi.re.kr}
\altaffiltext{2}{Korea University of Science and Technology, Daejeon, 34113, Republic of Korea}
\altaffiltext{3}{Institute of Space and Astronautical Science, Japan Aerospace Exploration Agency, Kanagawa, 252-5210, Japan}
\altaffiltext{4}{Genesia Corporation, Mitaka, Tokyo 181-0013, Japan}
\altaffiltext{5}{Kyung Hee University, Gyeonggi-do, 17104, Republic of Korea}

\begin{abstract}
The Multipurpose InfraRed Imaging System (MIRIS) performed the MIRIS Pa$\alpha$ Galactic Plane Survey (MIPAPS), which covers the entire Galactic plane within the latitude range of $-3\arcdeg \la b \la +3\arcdeg$ at Pa$\alpha$ (1.87 $\mu$m). We present the first result of the MIPAPS data extracted from the longitude range of $\ell = 96\arcdeg.5$--$116\arcdeg.3$, and demonstrate the data quality and scientific potential of the data by comparing them with H$\alpha$ maps obtained from the INT Photometric H$\alpha$ Survey (IPHAS) data. We newly identify 90 \ion{H}{2} region candidates in the {\it WISE} \ion{H}{2} region catalog as definite \ion{H}{2} regions by detecting the Pa$\alpha$ and/or H$\alpha$ recombination lines, out of which 53 \ion{H}{2} regions are detected at Pa$\alpha$. We also report the detection of additional 29 extended and 18 point-like sources at Pa$\alpha$. We estimate the $E(\bv)$ color excesses and the total Lyman continuum luminosities for \ion{H}{2} regions by combining the MIPAPS Pa$\alpha$ and IPHAS H$\alpha$ fluxes. The $E(\bv)$ values are found to be systematically lower than those estimated from point stars associated with \ion{H}{2} regions. Utilizing the MIPAPS Pa$\alpha$ and IPHAS H$\alpha$ images, we obtain an $E(\bv)$ map for the entire region of the \ion{H}{2} region Sh2-131 with an angular size of $\sim$2\arcdeg.5. The $E(\bv)$ map shows not only numerous high-extinction filamentary features but also negative $E(\bv)$ regions, indicating H$\alpha$ excess. The H$\alpha$ excess and the systematic underestimation of $E(\bv)$ are attributed to light scattered by dust.
\end{abstract}


\section{Introduction} \label{sec:intro}

Galactic \ion{H}{2} regions are important observational targets to study various astronomical phenomena in our Galaxy. Since massive stars, which produce the ultraviolet (UV) photons ionizing neutral hydrogens surrounding the stars, have short lifetimes, Galactic \ion{H}{2} regions are good indicators of the recent star-forming activity in our Galaxy. \citet{anderson14} presented a catalog of 8399 Galactic \ion{H}{2} regions using the {\it WISE} all-sky data, and claimed that it is the most complete catalog of Galactic \ion{H}{2} regions. However, they used an indirect method to identify \ion{H}{2} regions: searching the {\it WISE} 12 $\mu$m and 22 $\mu$m emissions, each of which originates from the polycyclic aromatic hydrocarbon (PAH) molecules and very small dust grains, respectively, which are heated by photons from the massive stars within the \ion{H}{2} regions \citep[e.g.,][]{li02}. Therefore, further studies to directly probe ionized hydrogen gas in the \ion{H}{2} region candidates are required to definitely confirm that they are indeed true \ion{H}{2} regions. There have been many radio continuum surveys along the Galactic plane \citep[e.g.,][]{becker94,condon98,bock99,taylor03,mcclure05,helfand06,stil06}. No single radio continuum survey covers the whole Galactic plane, but the combined coverage of those surveys does. \citet{paladini03} provided a radio catalog of 1442 Galactic \ion{H}{2} regions by collecting the previous radio data. Although the radio continuum observations are free from dust extinction, the radio continuum arises not only from thermal free-free emission produced in ionized hydrogen gas, but also from non-thermal radiation with other origins. Therefore, detecting hydrogen recombination lines is the best way to definitely identify \ion{H}{2} regions.

To date, the recombination lines from ionized hydrogen gas have been detected mostly by observing radio recombination lines (RRLs), near-infrared Br$\gamma$ 2.17 $\mu$m line, and optical H$\alpha$ line, because these lines are accessible in the ground-based observations. The most comprehensive RRL observations of Galactic \ion{H}{2} regions were performed with the Green Bank and Arecibo telescopes \citep{bania10,anderson11,bania12}. They detected RRLs from $\sim$500 \ion{H}{2} regions in the first Galactic quadrant and around the Galactic center. A large-scale Br$\gamma$ survey of the Galactic plane was carried out by \citet{kutyrev01}, but it covered only a limited longitude range ($\ell = 358\arcdeg$--$43\arcdeg$) of the plane. Since the RRLs and Br$\gamma$ lines are very faint, they do not seem to be efficient for large-scale survey. For example, the Br$\gamma$ intensity is about two orders of magnitude lower than the H$\alpha$ intensity, as shown in Table 14.2 of \citet{draine11}. The ground-based optical surveys of the H$\alpha$ line have been performed for more than half a century, and hence they are the most common methods for observing hydrogen recombination lines from \ion{H}{2} regions. Although an all-sky composite H$\alpha$ map \citep{finkbeiner03} is the most well known, the INT/WFC Photometric H$\alpha$ Survey (IPHAS) \citep{drew05} and the SuperCOSMOS H$\alpha$ Survey (SHS) \citep{parker05}, which cover the whole northern and southern Galactic plane, respectively, are more recent data with much higher spatial resolutions and sensitivities. However, a fundamental limitation of H$\alpha$ observations is the strong attenuation by dust, especially in the Galactic plane. Therefore, the detection of the H$\alpha$ line is not easy, not only for distant sources with a large amount of interstellar dust along the line of sight, but also for young \ion{H}{2} regions embedded within dense molecular clouds.

An alternative way to detect hydrogen recombination lines from \ion{H}{2} regions is Pa$\alpha$ 1.87 $\mu$m line observation. The Pa$\alpha$ line is still moderately strong in case B hydrogen recombination spectrum ($\sim$1/8.5 of H$\alpha$ intensity at a temperature of 10$^{4}$ K, as shown in Table 14.2 of \citet{draine11}), but suffers much less dust extinction than the H$\alpha$ line. Assuming the Galactic extinction curve of \citet{cardelli89} with $R_V$ = 3.1, the less-attenuated Pa$\alpha$ flux becomes larger than the more-attenuated H$\alpha$ flux when $E(\bv)$ $>$1.12. Therefore, the Pa$\alpha$ line observation enables us to probe deeper into dense molecular clouds and longer distances through the interstellar medium. However, the Pa$\alpha$ line is included in a wavelength range that is subject to heavy absorption by water molecules in Earth's atmosphere, and therefore on the ground, the Pa$\alpha$ line is not easily accessible. Nevertheless, there have been some Pa$\alpha$ observations on the ground \citep[e.g.,][]{jones80,lilly87,hines91,hill96,falcke98,komugi12,tanabe13}, whose observing targets were, however, limited to early-type stars or external galaxies. The first space-based Pa$\alpha$ observations were performed with the {\it Hubble Space Telescope} ({\it HST}) NICMOS camera. {\it HST} NICMOS has high spatial resolution and sensitivity, but a small field of view. This makes its observations mainly focus on small-sized targets, such as external galaxies \citep[e.g.,][]{boker99,alonso01,quillen01,scoville01,alonso02,diaz08,liu13} or small-sized objects in our Galaxy \citep[e.g.,][]{schultz99,smith00,scoville03,mills11}. In particular, \citet{wang10} and \citet{dong11} made a large-scale Pa$\alpha$ mosaic image with a size of $\sim$$39\times15$ arcmin$^{2}$ around the Galactic center. To obtain the image of this angular size, they combined {\it HST} NICMOS data taken from thousands of exposure fields.

The Multipurpose InfraRed Imaging System (MIRIS) was developed for a wide-field survey covering the whole Galactic plane with a moderate spatial resolution (the best case is $\sim$52 arcsec) \citep{han14}. The MIRIS Pa$\alpha$ Galactic Plane Survey (MIPAPS) was performed using two narrow-band filters near the Pa$\alpha$ line, and the Pa$\alpha$ image over the whole Galactic plane has been completed for the first time. In this paper, we present the first result of MIPAPS for a particular region in the longitude range of $\ell = 96\arcdeg.5$--$116\arcdeg.3$ (around Cepheus). This region, which lies just outside the first Galactic quadrant, includes many Pa$\alpha$ sources with intermediate brightness. The data in this region are free from an observational artifact due to filter-wheel position offset (this issue will be described in subsequent papers), and thus are appropriate for an initial analysis of the MIPAPS data. By comparing the MIPAPS Pa$\alpha$ data with the IPHAS H$\alpha$ data, we demonstrate the scientific potential of the MIPAPS data. We visually inspect the Pa$\alpha$ and H$\alpha$ data of \ion{H}{2} region candidates, and show that the MIPAPS Pa$\alpha$ data are able to identify many of them as true \ion{H}{2} regions. We also report new detections of Pa$\alpha$ extended and point-like sources in Cepheus. For some \ion{H}{2} regions, we estimate the $E(\bv)$ color excesses by aperture photometries of the Pa$\alpha$ and H$\alpha$ total fluxes, and compare the results with $E(\bv)$ obtained from point stars. We show that our photometric results can be used to constrain distances to \ion{H}{2} regions and the spectral types of their ionizing stars. Additionally, we produce an $E(\bv)$ map of an \ion{H}{2} region using the MIPAPS Pa$\alpha$ and IPHAS H$\alpha$ images, and present the radial profile of $E(\bv)$ for the \ion{H}{2} region.

\section{Observations and Data Reduction} \label{sec:data}

\subsection{MIRIS Pa$\alpha$ Observations} \label{subsec:miris}

MIRIS is the primary scientific payload of the third Korean Science and Technology Satellite in Korea, {\it STSAT-3}, which was launched in November 2013. The pixel size and field of view are 51.6 arcsec and $3\arcdeg.67\times3\arcdeg.67$, respectively. It has two broad-band filters, I (centered at $\sim$1.05 $\mu$m) and H (centered at $\sim$1.6 $\mu$m), and two narrow-band filters for the Pa$\alpha$ line (PAAL; centered at $\sim$1.875 $\mu$m) and the Pa$\alpha$ dual continuum (PAAC; centered at $\sim$1.84 $\mu$m and $\sim$1.91 $\mu$m). One of the main goals of MIRIS is to perform the Pa$\alpha$ Galactic plane survey using the two narrow-band filters PAAL and PAAC. From April 2014 until May 2015, the survey covered the entire Galactic plane within the latitude range of $-3\arcdeg \la b \la +3\arcdeg$. This contains 235 fields in total, with an average exposure of $\sim$20 minutes per filter. The details of the MIRIS instrument design are described in \citet{han14}, and its mission, on-orbit performance, and basic data processing pipeline will be addressed in subsequent papers. The entire data set went public on July, 2017 at the official MIRIS homepage\footnote{\url{http://miris.kasi.re.kr/miris/}}.

We extracted the pipeline-processed data for 14 fields in a longitude range of $\ell = 96\arcdeg.5$--$116\arcdeg.3$, which were obtained during 77 orbits (39 for PAAL, and 38 for PAAC). For each orbit, we adopted a median combined image from $\sim$180 frame images obtained during the orbit, which were provided by the MIRIS data processing pipeline. The background brightness of each orbit's image can increase due to the lunar light when its incident angle is less than 80$\arcdeg$. To remove the difference in the lunar backgrounds between the PAAL and PAAC images, we calculated a median background value for each orbit's image, and subtracted the constant background from all of the pixels. After that, we combined the images from every orbit using the Montage software\footnote{\url{http://montage.ipac.caltech.edu/}}, and obtained final mosaic images at the PAAL and PAAC bands for the Cepheus region.

\begin{figure}
\centering
\includegraphics[scale=0.7]{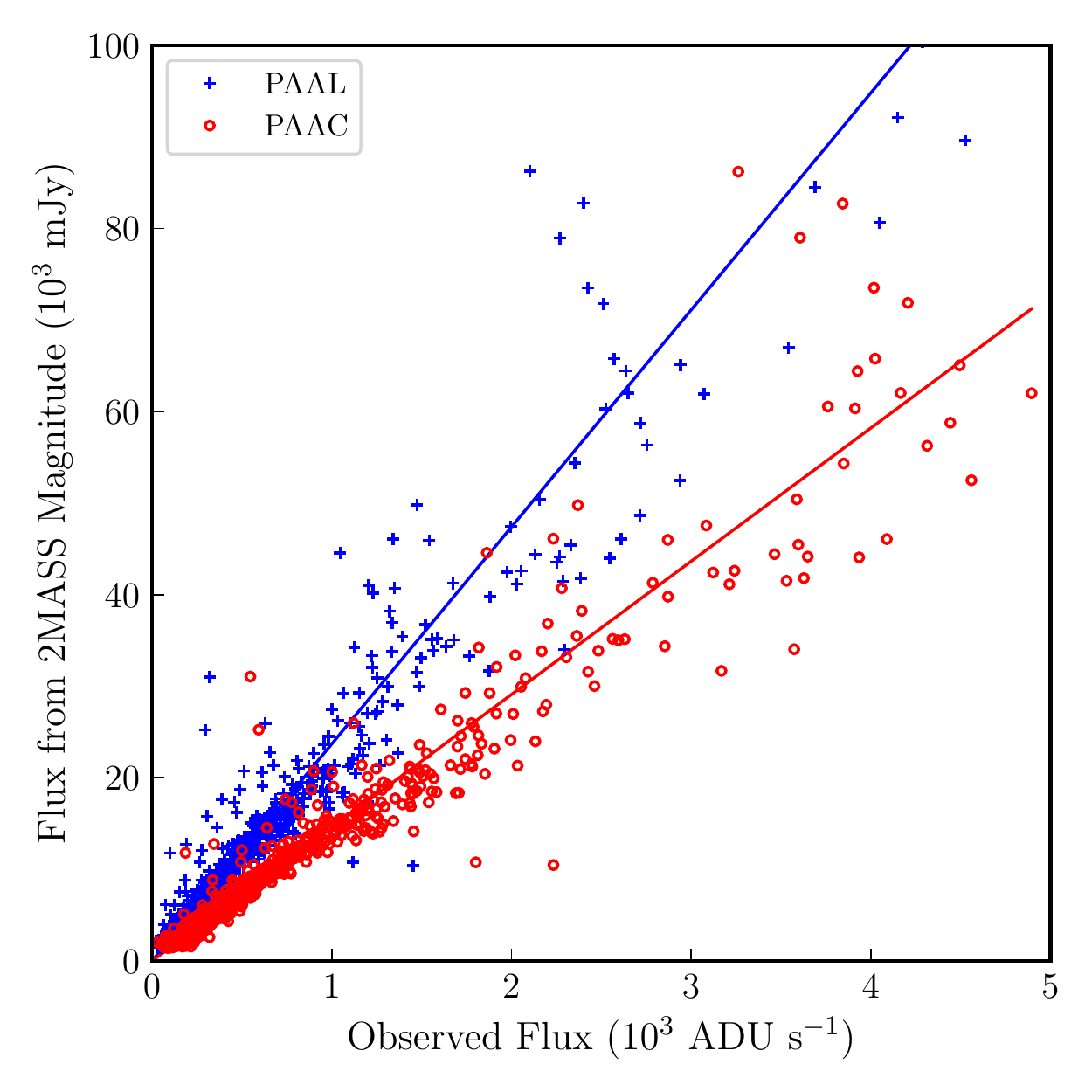}
\caption{MIRIS PAAL and PAAC flux calibration by comparison with the 2MASS magnitudes. The blue crosses and red circles indicate the fluxes calculated for the sources extracted from PAAL and PAAC mosaic images of the Cepheus region, respectively. The blue and red lines represent the best-fit lines, with slopes of 23.70 and 14.55 mJy (ADU s$^{-1}$)$^{-1}$, respectively.\label{fig:cal}}
\end{figure}

To perform the absolute calibration of the MIRIS flux at the PAAL and PAAC bands, we matched point sources detected in the PAAL and PAAC mosaic images with their counterparts in the 2MASS point-source catalog \citep{skrutskie06}. Using the Source Extractor package \citep{bertin96}, point sources were identified, and their fluxes in units of ADU s$^{-1}$ were measured in the PAAL and PAAC images. Here, ADU means Analog-to-Digital Unit \citep{han14}. We used the AUTO photometry (flux within a Kron-like elliptical aperture), and selected only the sources with the signal-to-noise ratio of $>$3. We also excluded too bright sources (29 at PAAL and 49 at PAAC) with the fluxes of $>$5000 ADU s$^{-1}$, because their large flux errors can cause significant uncertainty in estimating the calibration factors (to convert from ADU s$^{-1}$ to mJy). Applying the PAAL and PAAC transmission curves to the fluxes that were obtained by interpolating the 2MASS H and K$_{s}$ magnitudes of the point sources, the fluxes expected at PAAL and PAAC (in units of mJy) of the point sources were calculated. The MIRIS fluxes of a total of 1548 point sources for PAAL (blue crosses) and 1323 point sources for PAAC (red circles) are compared with those calculated from the 2MASS data in Figure \ref{fig:cal}. By the least-square linear fitting, we obtained the PAAL and PAAC calibration factors of 23.70 $\pm$ 0.14 and 14.55 $\pm$ 0.11 mJy (ADU s$^{-1}$)$^{-1}$, respectively.

\begin{figure*}
\centering
\includegraphics[scale=0.38]{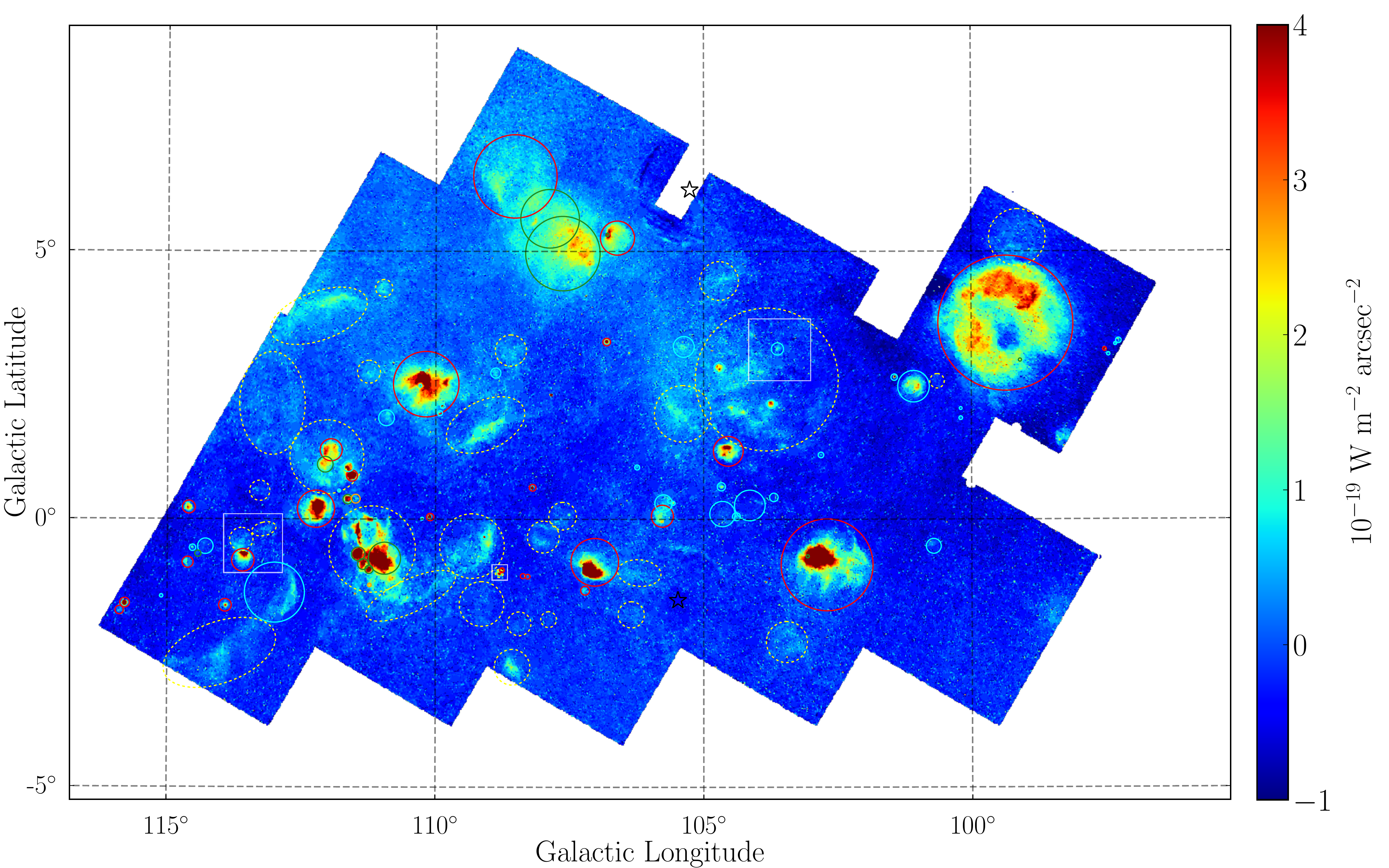}
\caption{Continuum-subtracted MIPAPS Pa$\alpha$ mosaic image of the Cepheus region in Galactic coordinate. The pixel size of the image is $\sim$52 arcsec. The solid circles indicate 80 Pa$\alpha$ sources corresponding to {\it WISE} \ion{H}{2} region sources listed in Tables \ref{table:wk}--\ref{table:wr}: Red, cyan, green, and orange colors represent the Pa$\alpha$ sources corresponding to 27 ``Known'', 39 ``Candidate'', 12 ``Group'', and 2 ``Radio Quiet'' {\it WISE} sources, respectively. The yellow dotted circles and ellipses indicate the other 29 Pa$\alpha$ extended sources with no counterparts in the {\it WISE} \ion{H}{2} region catalog; they are listed in Table \ref{table:mpe}. Three rectangles designate the regions selected as the examples of visual inspection in Figure \ref{fig:example}. Two star symbols indicate the positions of two infrared-bright stars that cause some artifact features that are centered on them.\label{fig:paa}}
\end{figure*}

Subtraction of the calibrated PAAC image from the calibrated PAAL image results in the continuum-subtracted Pa$\alpha$ image. Because PAAC is a dual continuum filter, the dust reddening effect on the PAAC flux is not significantly different from that on the PAAL flux. Applying the extinction curve of \citet{cardelli89} with $R_V$ = 3.1 to the PAAL and PAAC fluxes, the difference in reddening effect between the two bands is calculated to be at most $\sim$0.8\%, even when $E(\bv)$ = 10. Although the difference in dust reddening is negligible, a straightforward subtraction of the PAAC image from the PAAL image leaves strong residual patterns around bright stars. This is mainly due to a significant difference between the point spread functions (PSFs) at PAAL and PAAC. The method of using a common PSF and image convolution worsens the spatial resolution of images, and was found to still leave stellar residuals. Thus, we decided to mask out the residuals around the 2MASS point sources with H and K$_{s}$ magnitudes of $\leqq$8. We masked out a circular area around each bright star, and filled it with a median value of the neighboring annulus with the width of the masking radius. Radii of masking areas were empirically determined to be roughly proportional to the 2MASS magnitude of the star. Figure \ref{fig:paa} shows the continuum-subtracted Pa$\alpha$ image of the Cepheus region. The strength of the Pa$\alpha$ emission line is expressed by the in-band flux unit (W m$^{-2}$) integrated over the PAAL transmission curve. We note that there are still a few remaining artifacts due to two infrared-bright stars, HD 214665 at ($\ell$, $b$) = ($105\arcdeg.48, -1\arcdeg.51$) and HD 209772 at ($\ell$, $b$) = ($105\arcdeg.26, 6\arcdeg.15$). Streaks or arc features, around ($\ell$, $b$) = ($105\arcdeg.30, -0\arcdeg.58$), ($105\arcdeg.70, 5\arcdeg.47$), and ($106\arcdeg.10, 6\arcdeg.48$), centered on these stars are caused by the PSF difference between the PAAL and PAAC bands.

\begin{figure*}
\centering
\includegraphics[scale=0.38]{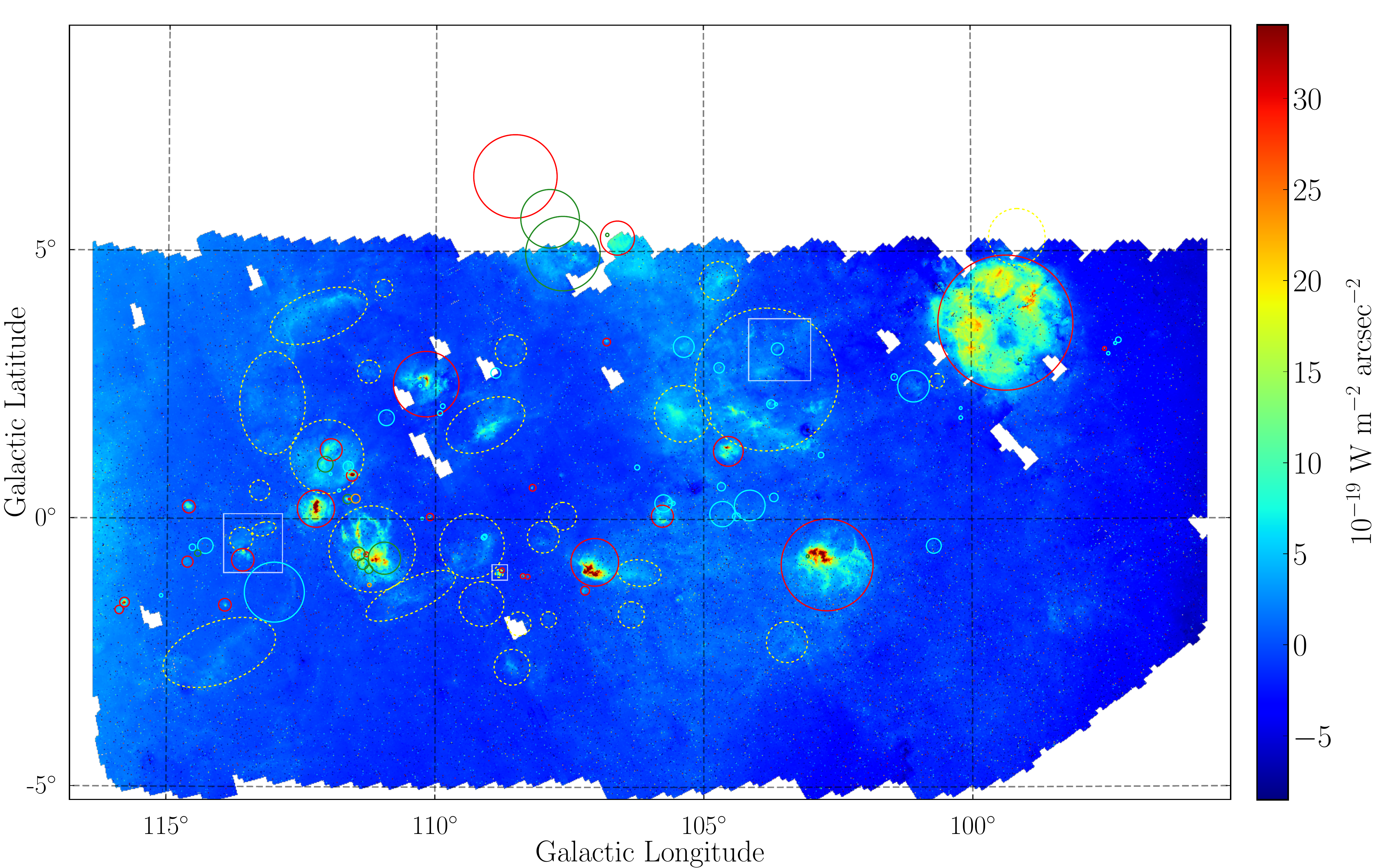}
\caption{Continuum-subtracted IPHAS H$\alpha$ mosaic image of the Cepheus region in Galactic coordinate. This image was plotted with a pixel size of $\sim$10 arcsec, although the original mosaic data have a bin size of $\sim$5 arcsec. The color-scale range has been adjusted by multiplying that of Figure \ref{fig:paa} (-1 to 4 $\times$ 10$^{-19}$ W m$^{-2}$ arcsec$^{-2}$) by the intrinsic H$\alpha$ to Pa$\alpha$ ratio ($\sim$8.5). All of the overlaid circles, ellipses, and rectangles are the same as in Figures \ref{fig:paa}.\label{fig:ha}}
\end{figure*}

\subsection{IPHAS H$\alpha$ Survey data} \label{subsec:iphas}

To compare the MIPAPS Pa$\alpha$ data with the existing H$\alpha$ data, we used the recently calibrated IPHAS data \citep{barentsen14}. The continuum-subtracted H$\alpha$ image was obtained combining the narrow-band H$\alpha$ and broad-band $r$ filter data. For each filter, we selected a total of 7840 IPHAS field data in the Cepheus region, which satisfy all of the IPHAS quality control criteria \citep{barentsen14}. We masked out bad or low-confidence pixels, by excluding the pixels with the ``confidence'' value of $<$70 in the ``confidence'' maps provided for each field data \citep{gonzalez08}. Additionally, we discarded all pixels in the first five columns along the longer side of each IPHAS CCD. These pixels were found to have quite different values from those of other pixels in almost all field data, which has not been reflected in the confidence maps. We subtracted the median of sky background from each field image to reduce the difference in the sky background between the H$\alpha$ and $r$ filter images. Since the IPHAS pixel size (0.33 arcsec) is too small to make such a large, single mosaic image for the whole Cepheus region, each field image was binned by 15 pixels $\times$ 15 pixels that corresponds to an angular size of $\sim$5 arcsec. Then, all field images at each band (H$\alpha$ and $r$) were combined into a final mosaic image for the Cepheus region using the Montage software.

The pixel values were converted into the calibrated flux units (in W m$^{-2}$ \AA{$^{-1}$}) by substituting the value of photometric zero-point presented in each image header into Equation (11) of \citet{barentsen14}, and using zero magnitudes of the IPHAS H$\alpha$ and $r$ filters obtained from a web site for filter profile service\footnote{\url{http://svo2.cab.inta-csic.es/theory/fps/}}. Direct subtraction of the $r$ filter image from the H$\alpha$ filter image underestimates the continuum-subtracted H$\alpha$ emission, because the wavelength range of the narrow-band H$\alpha$ filter is included in that of the broad-band $r$ filter. Therefore, we multiplied the resulting image by a correction factor of $(\Sigma T_{H\alpha} \times \Sigma T_r) / (\Sigma T_r - \Sigma T_{H\alpha})$ calculated from the IPHAS H$\alpha$ and $r$ filter transmission profiles (available from the above web site). Here, $\Sigma T_{H\alpha}$ and $\Sigma T_r$ are the integrals of the H$\alpha$ and $r$ filter transmission curves over wavelength, respectively, and thus multiplying the correction factor expresses the strength of the H$\alpha$ emission line in the in-band flux unit (in W m$^{-2}$). As in the continuum-subtracted Pa$\alpha$ image, the continuum-subtracted H$\alpha$ image also shows strong residual features around most of the bright stars, due to the PSF difference between the H$\alpha$ and $r$ bands. Moreover, the IPHAS data with a higher spatial resolution and sensitivity contain a quite large number of optical point stars in the whole Cepheus region. Therefore, we masked out only the residuals around individual interesting sources while inspecting close-up images or performing photometry for them, as will be mentioned in Section \ref{sec:results}. To determine masking positions, we used the IPHAS source catalog \citep{barentsen14}. We selected the point sources with $r$-band magnitude (``r'' column) of $\leqq$16 and point-source probability (``pStar'' column) of $\geqq$0.5 in the catalog. The restriction condition of ``pStar'' $\geqq$0.5 prevented real diffuse sources from being masked out. We masked out a circular area around each point source. The masking radius was determined to be roughly proportional to the $r$-band magnitude of the source. The masked circular area was filled with a median value of the neighboring annulus with the width of the masking radius. However, there still remained many point-like sources with residuals even after performing the above procedures. Thus, we additionally masked out these residuals by visually inspecting the residuals in close-up H$\alpha$ images for individual sources. We checked the close-up images several times, especially for the sources selected to perform photometry, until the photometric results were not significantly affected by the remaining residuals. Figure \ref{fig:ha} shows the continuum-subtracted H$\alpha$ mosaic image of the whole Cepheus region. The color scale in the figure was adjusted to have the same color scale as in Figure \ref{fig:paa} ($-$1 to $4\times 10^{-19}$ W m$^{-2}$ arcsec$^{-2}$) after multiplying the intrinsic H$\alpha$ to Pa$\alpha$ ratio ($\sim$8.5), which makes it easy to recognize high-extinction regions from a simple comparison of the two line images.

\section{Results} \label{sec:results}

\subsection{WISE \ion{H}{2} region sources} \label{subsec:wise_sources}

\begin{deluxetable*}{ccccccccccc}
\tablecaption{Summary of Visual Inspection for {\it WISE} \ion{H}{2} Region Sources\label{table:vis}}
\tablewidth{0pt}
\tabletypesize{\footnotesize}
\tablehead{
\colhead{Category} & \colhead{Number of Sources} & \multicolumn{3}{c}{MIPAPS Pa$\alpha$ Detection} &&
\multicolumn{4}{c}{IPHAS H$\alpha$ Detection} & \colhead{Pa$\alpha$ and/or H$\alpha$ Detection} \\
\cline{3-5} \cline{7-10}
\colhead{} & \colhead{} & \multicolumn{2}{c}{Yes} & \colhead{No} &&
\multicolumn{2}{c}{Yes} & \colhead{No} & \colhead{Not Observed} & \colhead{}
}
\startdata
``Known'' & 31 & 27 & (87.1\%) & 4 && 27 & (87.1\%) & 3 & 1 & 28 \\
``Candidate'' & 71 & 39 & (54.9\%) & 32 && 53 & (74.6\%) & 18 & 0 & 54 \\
``Group'' & 18 & 12 & (66.7\%) & 6 && 16 & (88.9\%) & 0 & 2 & 18 \\
``Radio Quiet'' & 92 & 2 & (2.2\%) & 90 && 18 & (19.6\%) & 70 & 4 & 18 \\
\hline
Total & 212 & 80 & (37.7\%) & 132 && 114 & (53.8\%) & 91 & 7 & 118 \\
\enddata
\tablecomments{All numbers denote the numbers of sources except for those in parentheses which are detection percentages. In the last column, 54 ``Candidate'', 18 ``Group'', and 18 ``Radio Quiet'' sources are newly identified as definite \ion{H}{2} regions by detecting the Pa$\alpha$ and/or H$\alpha$ recombination lines.
}
\end{deluxetable*}

\begin{figure*}
\centering
\includegraphics[scale=0.38]{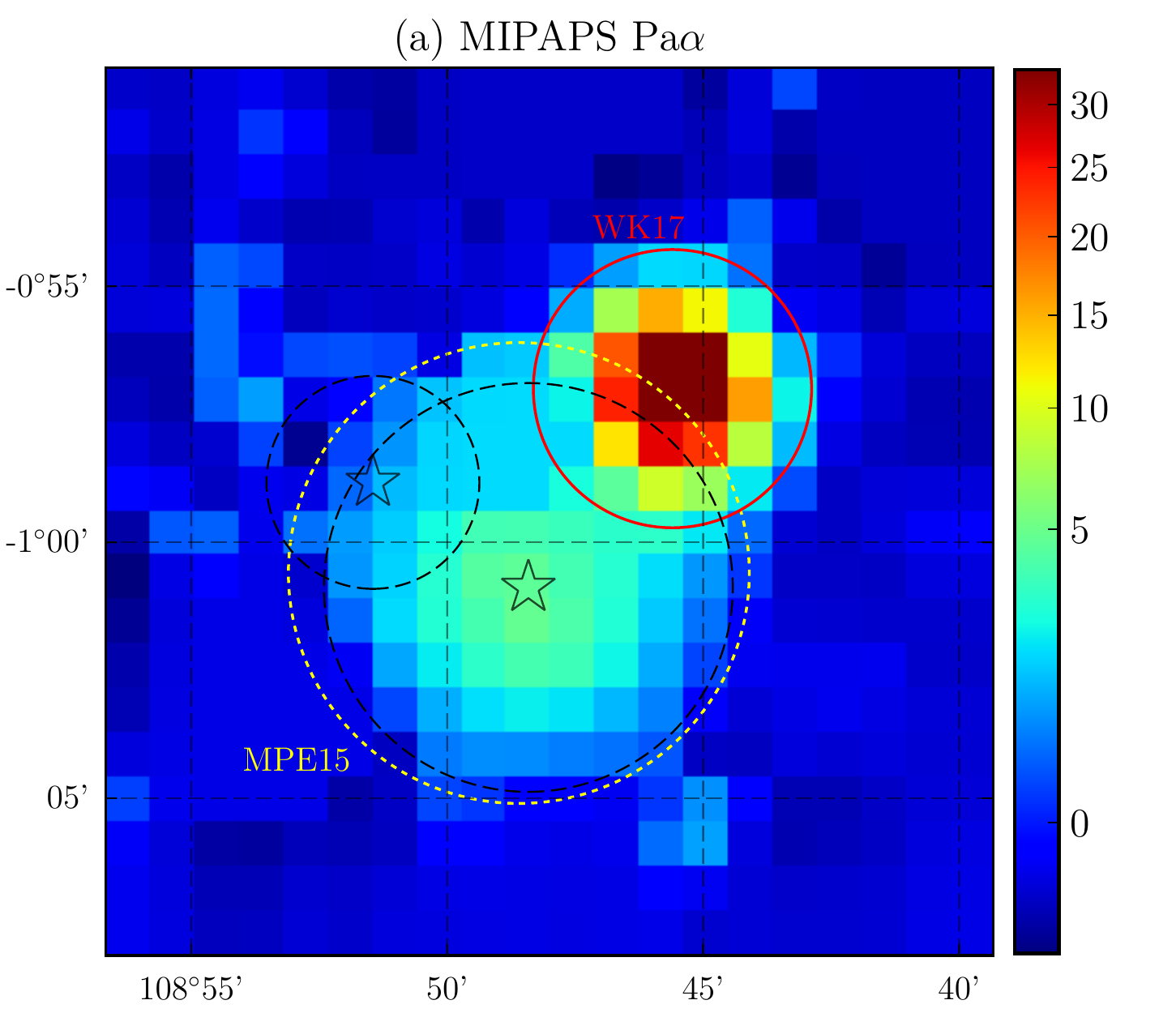}\includegraphics[scale=0.38]{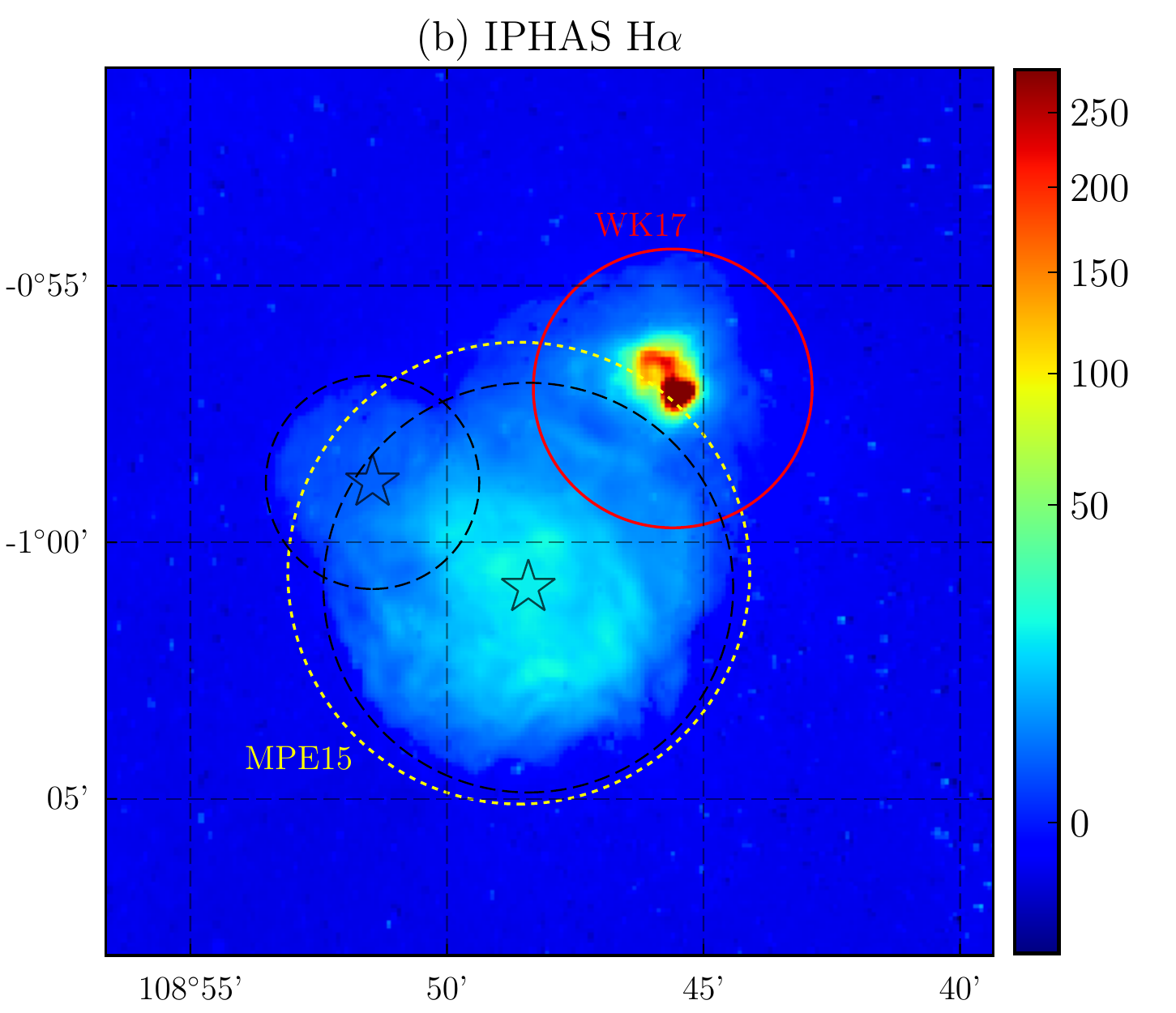}\includegraphics[scale=0.38]{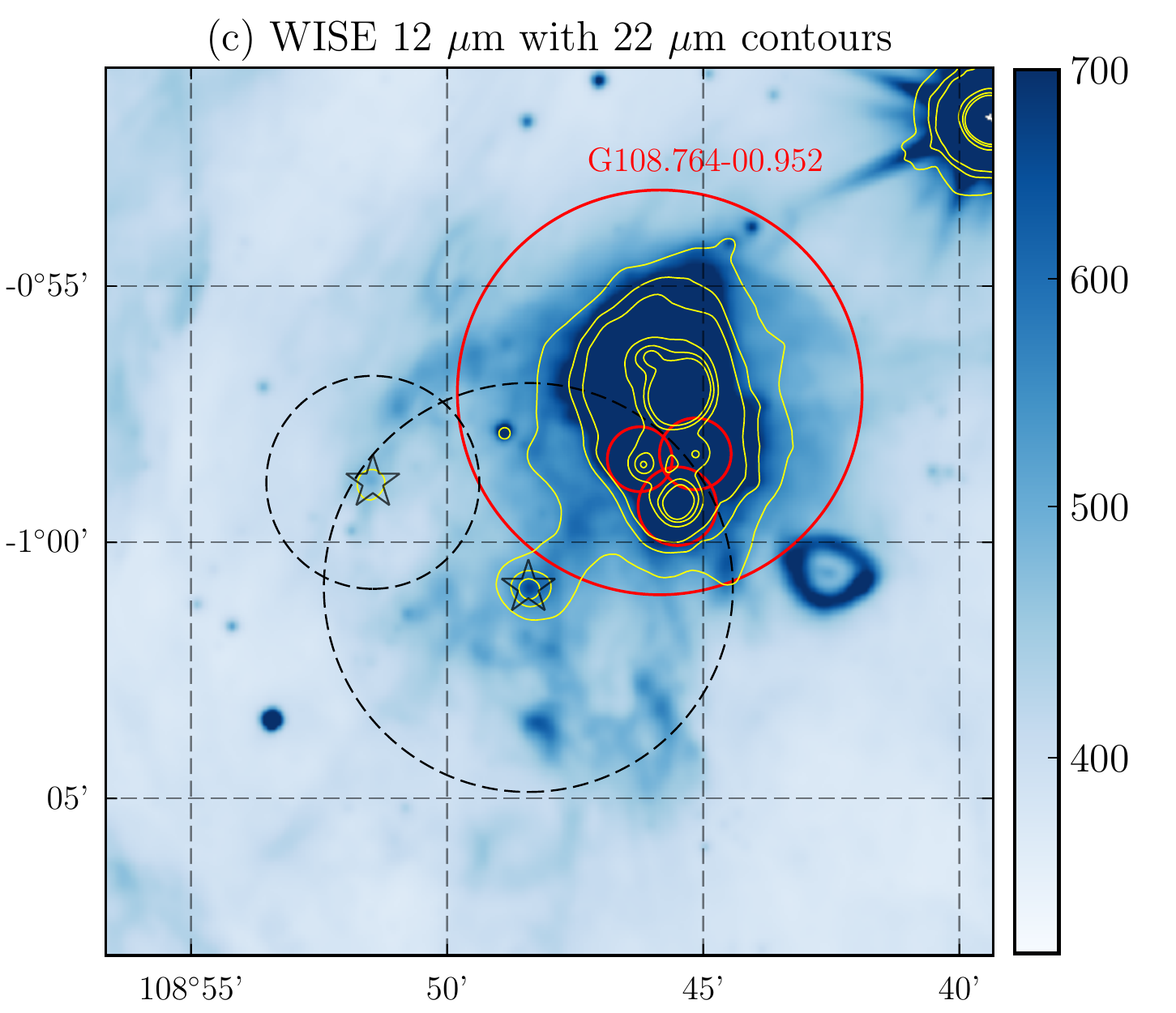}
\includegraphics[scale=0.38]{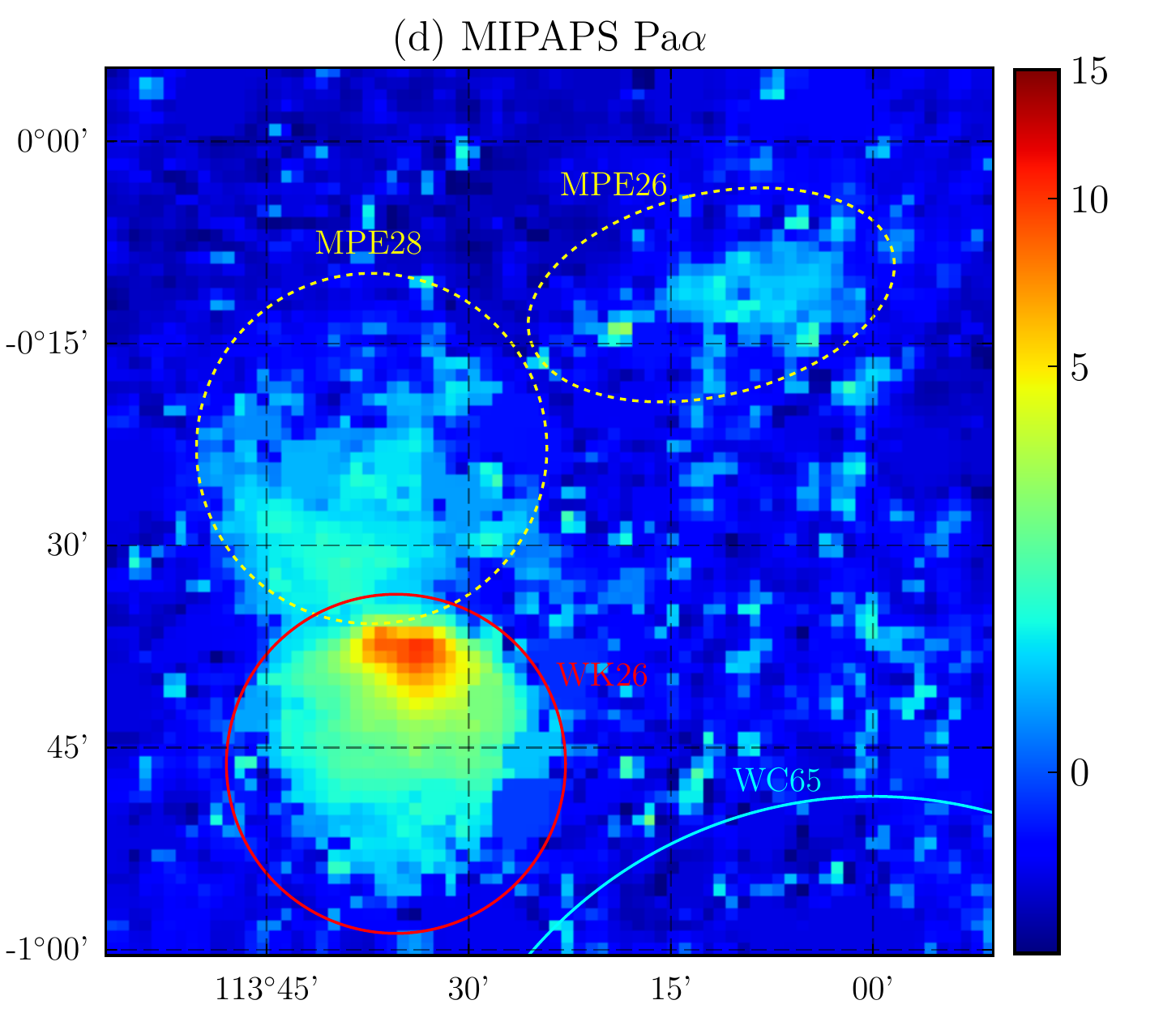}\includegraphics[scale=0.38]{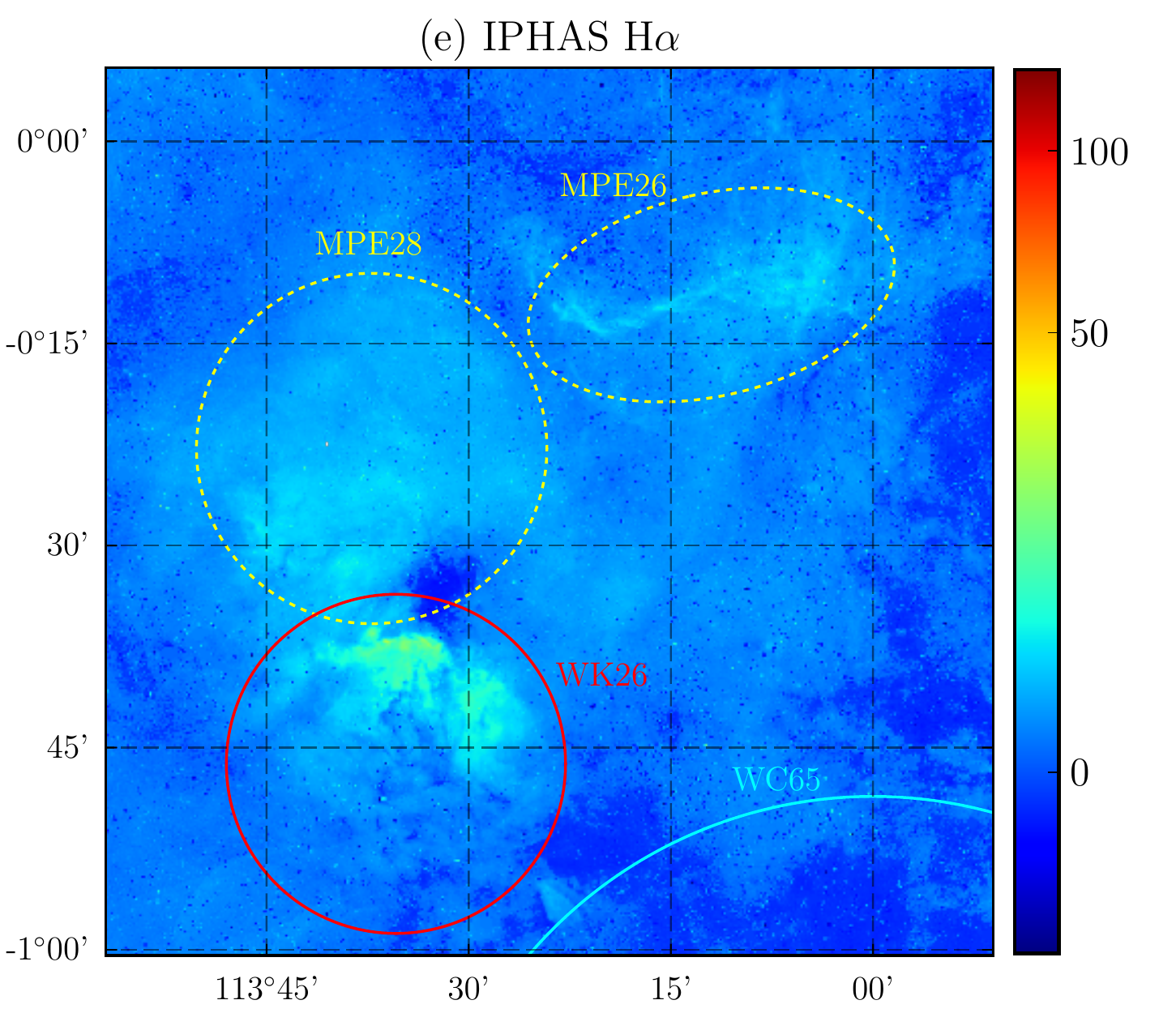}\includegraphics[scale=0.38]{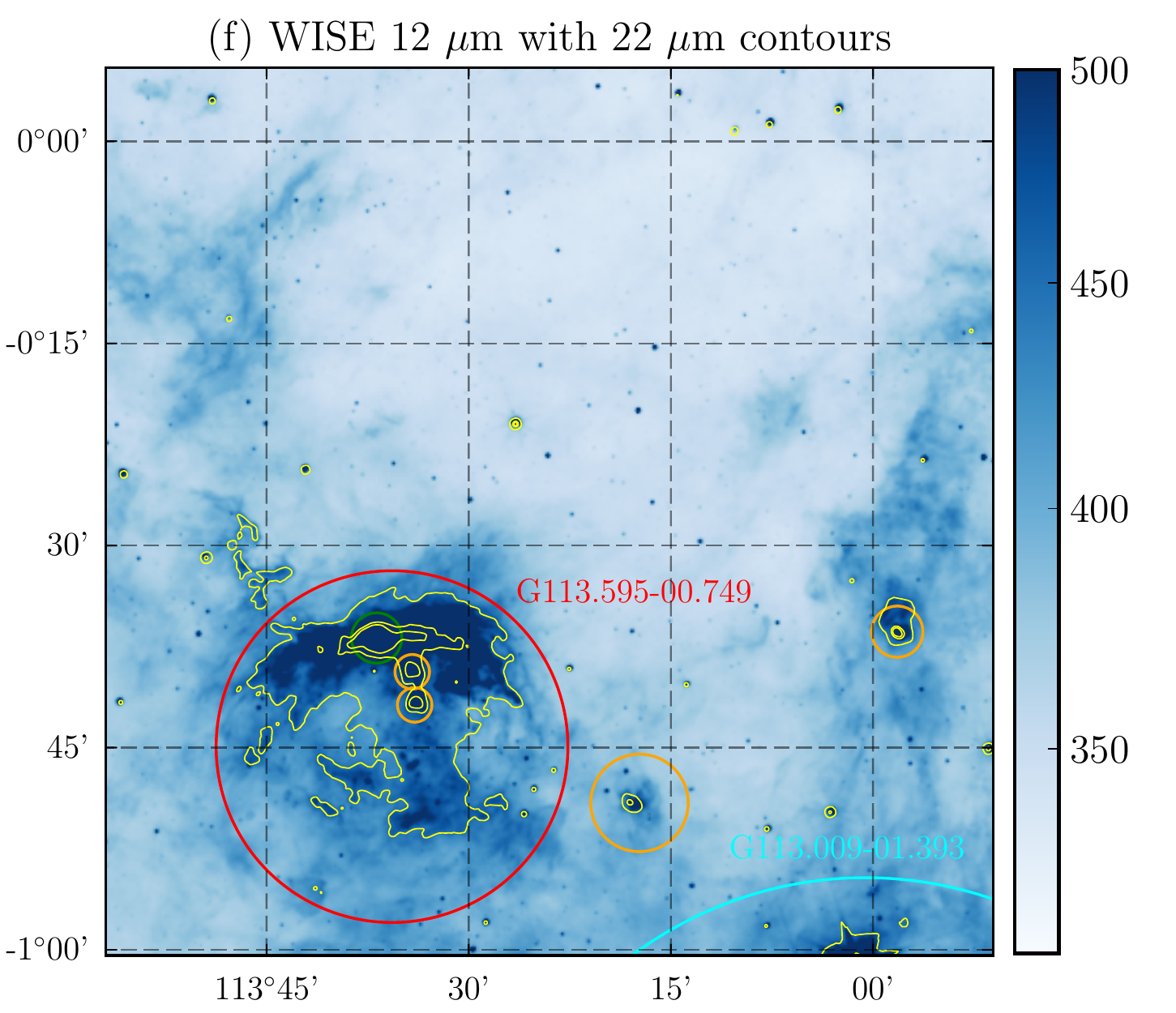}
\includegraphics[scale=0.38]{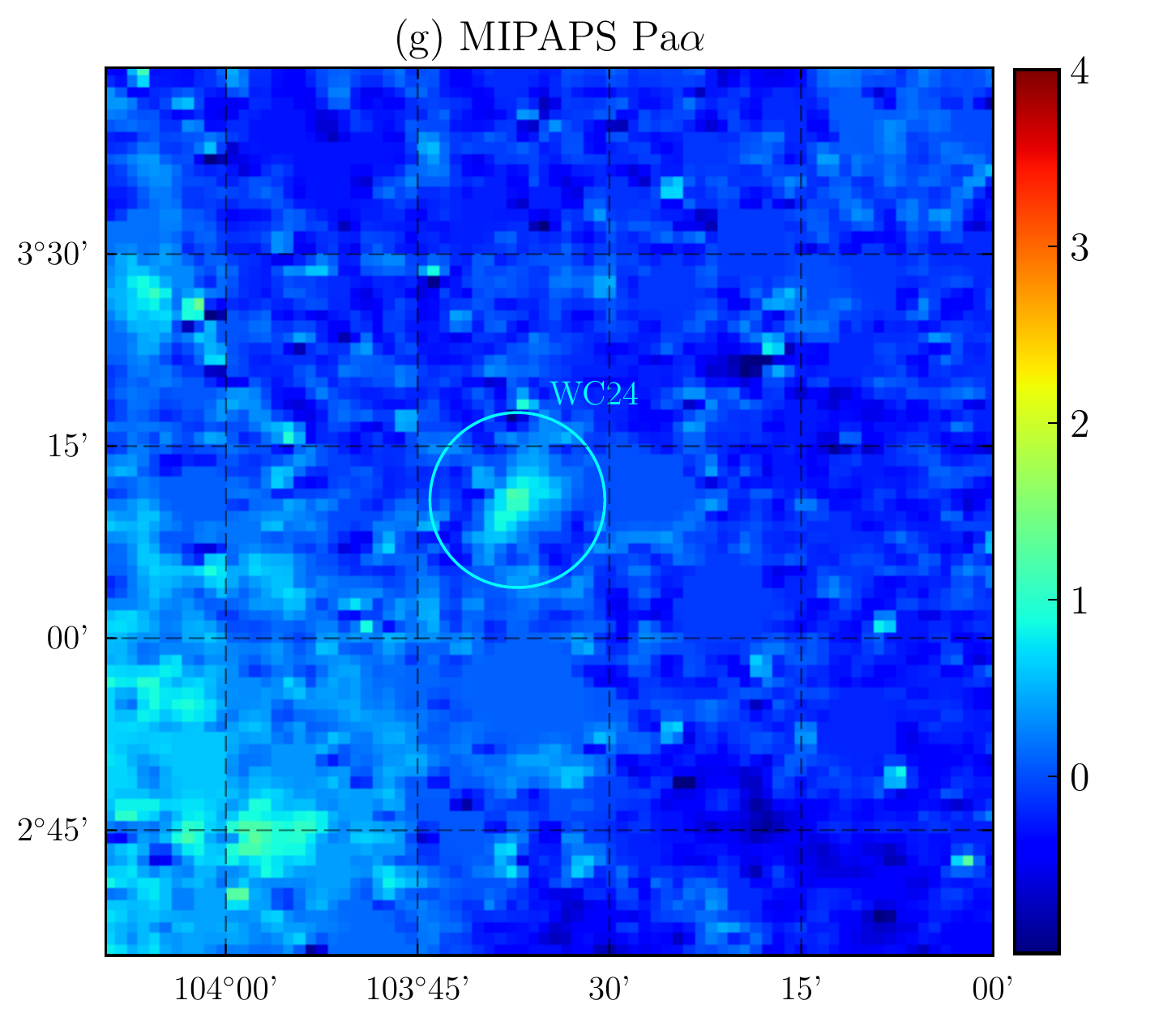}\includegraphics[scale=0.38]{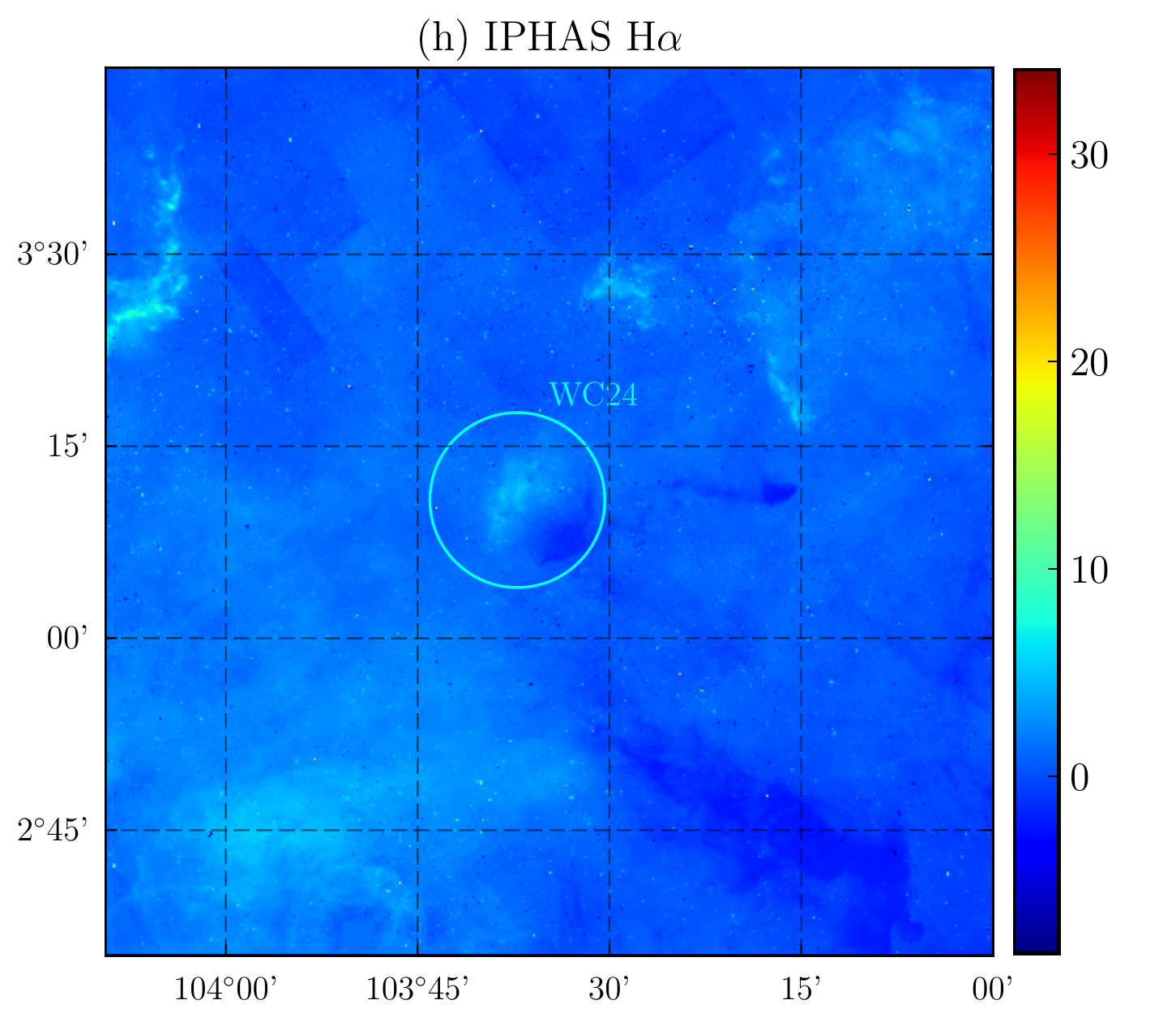}\includegraphics[scale=0.38]{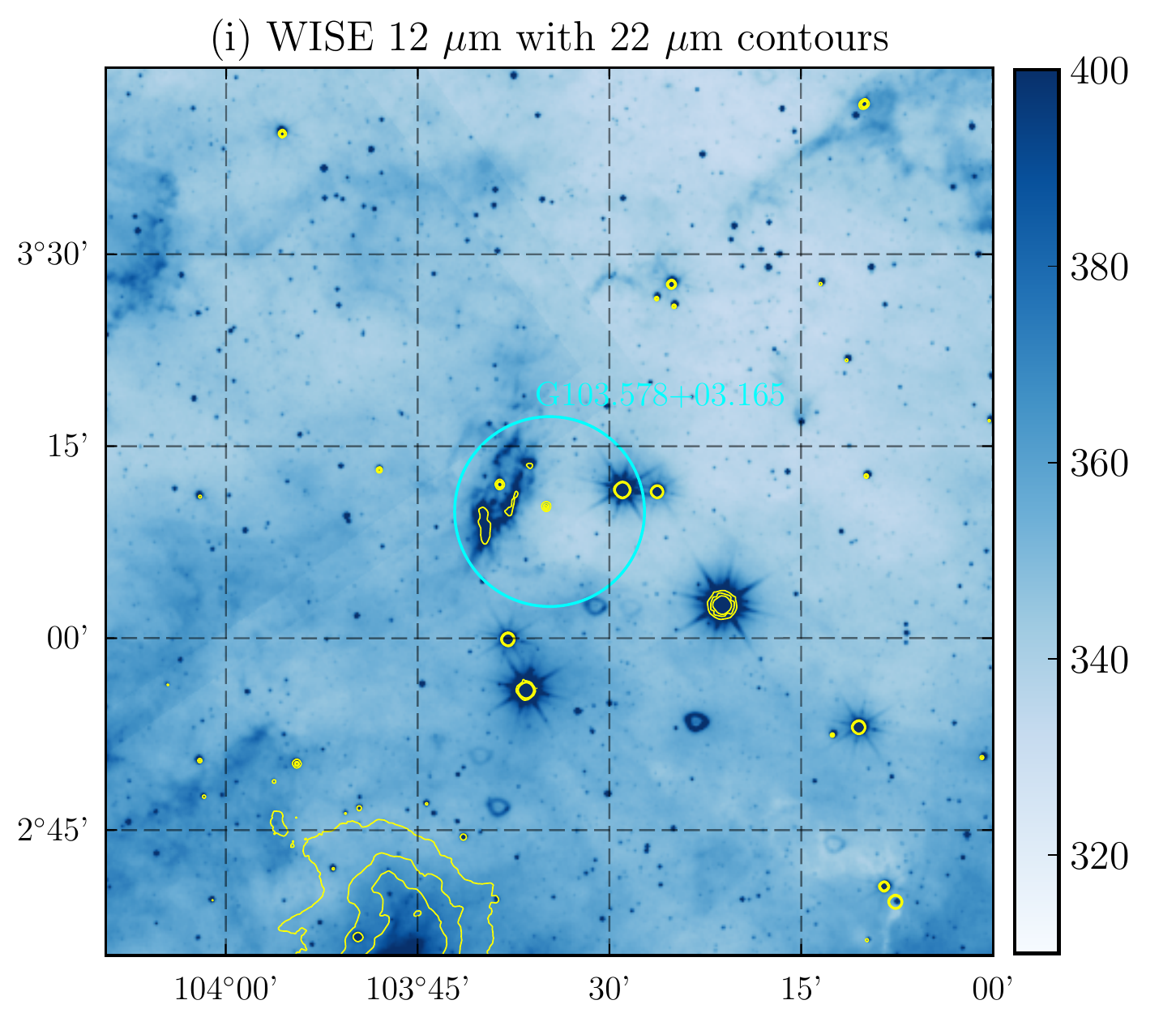}
\caption{MIPAPS Pa$\alpha$ (left column), IPHAS H$\alpha$ (middle), and {\it WISE} 12 $\mu$m (with 22 $\mu$m contours) (right) images of three representative regions for visual inspection. All coordinate systems are Galactic coordinates, and the pixel sizes of the Pa$\alpha$, H$\alpha$, and {\it WISE} 12 $\mu$m images are $\sim$52, $\sim$5, and $\sim$1.4 arcsec, respectively. The units of the color bars for Pa$\alpha$ and H$\alpha$ are all 10$^{-19}$ W m$^{-2}$ arcsec$^{-2}$, and the color-scale ranges have been adjusted according to the intrinsic H$\alpha$ to Pa$\alpha$ ratio ($\sim$8.5). The units of the color bars for {\it WISE} 12 $\mu$m are DN (Digital Number), and the levels of {\it WISE} 22 $\mu$m contours are 105, 115, 125, 225, 325, and 425 DN in (c), 97, 117, and 137 DN in (f), and 97, 100, and 103 DN in (i). All colored circles and ellipses in the Pa$\alpha$ and H$\alpha$ images are the same as in Figures \ref{fig:paa} and \ref{fig:ha}, and their corresponding source names are from the ``ID'' columns of Tables \ref{table:wk}--\ref{table:wr} and \ref{table:mpe}. All colored circles in the {\it WISE} 12 $\mu$m images indicate the {\it WISE} \ion{H}{2} region sources, of which the central positions and sizes come from the {\it WISE} \ion{H}{2} region catalog \citep{anderson14}. {\it WISE} names for some principal sources are shown together. In each image of the first region (top row), two star symbols indicate two ionizing stars of Sh2-153 presented by \citet{foster15} (see Table \ref{table:dis}), and two black dashed circles are centered on the two stars.\label{fig:example}}
\end{figure*}

\citet{anderson14} made a comprehensive catalog of Galactic \ion{H}{2} regions using the {\it WISE} 12 $\mu$m and 22 $\mu$m data. They identified potential \ion{H}{2} region candidates based on the characteristic mid-infrared (MIR) morphology of \ion{H}{2} regions: the 12 $\mu$m emission (from PAH molecules) surrounds the 22 $\mu$m emission (from heated dust grains). They also searched for counterparts at radio continuum, RRL, and H$\alpha$ of their candidates to confirm them as true \ion{H}{2} regions. Counterparts at the RRL or H$\alpha$ line have been found only for $\sim$18\% out of 8399 candidates, which were classified as ``Known'' sources. About 24\% of the candidates, classified as ``Candidate'' sources, were found to have only radio continuum counterparts. About 8\% were found to be spatially associated with known \ion{H}{2} region complexes, and were thus classified as ``Group'' sources. About 49\% of the candidates, classified as ``Radio Quiet'' sources, were found to have no counterparts at any of the radio continuum, RRL, and H$\alpha$ at the sensitivity limits of existing surveys. The Cepheus region investigated in this study contains a total of 212 sources (31 ``Known'', 71 ``Candidate'', 18 ``Group'', and 92 ``Radio Quiet'' sources), which are listed in Tables \ref{table:wk}--\ref{table:wr}.

For individual {\it WISE} \ion{H}{2} region sources, we visually compared the continuum-subtracted MIPAPS Pa$\alpha$ and IPHAS H$\alpha$ images with the {\it WISE} 12 $\mu$m and 22 $\mu$m images to find whether they show clear morphological coincidence. The {\it WISE} data were obtained from the NASA/IPAC Infrared Science Archive\footnote{\url{http://irsa.ipac.caltech.edu/applications/wise/}}. Tables \ref{table:wk}--\ref{table:wr} present the results of the visual inspection for the 212 {\it WISE} \ion{H}{2} region sources, of which statistics are summarized in Table \ref{table:vis}. Figure \ref{fig:example} shows the examples of three representative regions. Table \ref{table:wk} shows that the Pa$\alpha$ and H$\alpha$ counterparts were found for all of the ``Known'' sources, except for 5 sources. Out of them, two examples, WK17 and WK26, are shown in Figures \ref{fig:example}(a)--\ref{fig:example}(c) and Figures \ref{fig:example}(d)--\ref{fig:example}(f), respectively. Three small ``Known'' sources (WK15, WK16, and WK18; indicated by three small red circles in Figure \ref{fig:example}(c)), show no clear counterparts at Pa$\alpha$ and H$\alpha$, which is likely due to overlapping with two bright sources, WK17 and MPE15, as shown in Figures \ref{fig:example}(a) and \ref{fig:example}(b). WK02 also overlaps with another bigger source, and hence could not be clearly identified in the Pa$\alpha$ image, but it is distinguishable in the H$\alpha$ image. WK14, of which the Pa$\alpha$ counterpart was found, is located outside the IPHAS survey area. As shown in Tables \ref{table:wc} and \ref{table:vis}, 39 and 53 sources out of 71 ``Candidate'' sources were confirmed, respectively, at Pa$\alpha$ and H$\alpha$. Figures \ref{fig:example}(g)--\ref{fig:example}(i) display one of them, WC24. A feature that shows similar shape at Pa$\alpha$ and H$\alpha$ seems to be spatially related to {\it WISE} bright features aligned along the east side of G103.578+03.165. Fifteen sources that were found only in the H$\alpha$ images have small angular sizes of $<$160 arcsec, and some of them are indistinguishable from other bright sources or stellar residuals in the Pa$\alpha$ images. On the other hand, WC55 shows a Pa$\alpha$ feature separated from an adjacent bright source, but it has no clear counterpart in the H$\alpha$ image. Tables \ref{table:wg} and \ref{table:vis} show that we found that Pa$\alpha$ emissions from 12 of the ``Group'' sources are distinguishable from spatially-associated \ion{H}{2} region complexes. The other 6 ``Group'' sources, which are invisible at Pa$\alpha$, were found in the IPHAS H$\alpha$ images with higher resolution. In Table \ref{table:wr}, most of the ``Radio Quiet'' sources are small in angular size, so we detected only two sources at Pa$\alpha$. However, additional 16 sources were found in the IPHAS H$\alpha$ images. For all of the sources that were detected at Pa$\alpha$, we assigned MIPAPS names according to the central positions of the individual Pa$\alpha$ features, and presented radius values that approximate their angular extent. Tables \ref{table:wk}--\ref{table:wr} list them, while Figures \ref{fig:paa} and \ref{fig:ha} denote the solid circles with their central positions and radii.

We carried out aperture photometry for the {\it WISE} \ion{H}{2} region sources detected at Pa$\alpha$ to measure the Pa$\alpha$ fluxes. The sources that largely overlap with stellar residuals or other bright sources were excluded, because their precise photometry could not be obtained. For the sources that were also detected at H$\alpha$ and were fully covered by the IPHAS observation, we performed aperture photometry of the H$\alpha$ fluxes as well. For both Pa$\alpha$ and H$\alpha$ aperture photometries, we adopted circular apertures with the MIPAPS coordinates and radii given in Tables \ref{table:wk}--\ref{table:wr}. Therefore, the same aperture was used for both Pa$\alpha$ and H$\alpha$ photometries of each individual source. Since we determined source radii to be large enough that each aperture fully covered the individual source, the calculated fluxes can be considered as the total fluxes of the sources. Each background annulus was defined with a thickness of twice the aperture radius, and a median in the annulus was taken as the background value. The 1$\sigma$ uncertainty of flux photometry was also defined as a standard deviation in the background annulus. By comparing the observed Pa$\alpha$-to-H$\alpha$ total flux ratio with the value predicted in the theory of radiative recombination of hydrogen atom, we can obtain the $E(\bv)$ color excess averaged over each \ion{H}{2} region. We assumed the case B condition at a temperature of 10$^{4}$ K, as in Table 14.2 of \citet{draine11}. The differential extinction $A_{Pa\alpha}-A_{H\alpha}$ was converted into $E(\bv)$, applying the extinction curve of \citet{cardelli89} with $R_V$ = 3.1. Tables \ref{table:wk}--\ref{table:wr} give the resulting flux photometry and $E(\bv)$, together with the 1$\sigma$ uncertainties. Because the H$\alpha$ fluxes of WC65 and WC71 were calculated to be negative values, we present only their Pa$\alpha$ fluxes. Hereafter, the $E(\bv)$ color excess calculated in this way will be called Pa$\alpha$-H$\alpha$ $E(\bv)$, if needed to discriminate from those calculated using other methods.

\subsection{MIPAPS Pa$\alpha$ sources} \label{subsec:mipaps_sources}

In Figure \ref{fig:paa}, we detected Pa$\alpha$ sources that are not included in the {\it WISE} \ion{H}{2} region catalog. We found 29 extended and 18 point-like sources, which are listed in Tables \ref{table:mpe} and \ref{table:mpp}, together with assigned MIPAPS names. MIPAPS names are based on the central positions of the individual Pa$\alpha$ features, and the spatial extents of extended sources are approximated by circles or ellipses with the radii and position angles given in Table \ref{table:mpe}. The circles and ellipses are denoted by yellow dotted lines in Figures \ref{fig:paa} and \ref{fig:ha}. Using the SIMBAD database, we listed known \ion{H}{2} regions or objects corresponding to the Pa$\alpha$ extended and point-like sources in Tables \ref{table:mpe} and \ref{table:mpp}. We found 16 of the Pa$\alpha$ extended sources to have \ion{H}{2} region entries in the SIMBAD database, whereas the remaining 13 sources have no known \ion{H}{2} region counterparts, even in SIMBAD. All of the 18 Pa$\alpha$ point-like sources were found to be associated with emission-line stars (including Wolf-Rayet and Herbig Ae/Be stars) or planetary nebulae.

Table \ref{table:mpe} shows the IPHAS H$\alpha$ counterparts that were found for all of the 29 Pa$\alpha$ extended sources. Even the 13 Pa$\alpha$ extended sources that have no corresponding known \ion{H}{2} regions show morphological coincidence between Pa$\alpha$ and H$\alpha$. Figures \ref{fig:example}(a)--\ref{fig:example}(f) show three examples (MPE15, MPE26, and MPE28). MPE15 corresponds to Sh2-153 in SIMBAD. Two star symbols denoted in Figures \ref{fig:example}(a)--\ref{fig:example}(c) indicate the positions of two ionizing stars of Sh2-153 presented by \citet{foster15}. The H$\alpha$ morphology of MPE15 clearly shows two circular features centered on the ionizing stars, as denoted by dashed circles in the figures. The Pa$\alpha$ image shows similar morphology to the H$\alpha$ image, although the smaller feature is not clear. The two ionizing stars are also spatially coincident with two {\it WISE} 22 $\mu$m local peaks, as shown by the yellow contours in Figure \ref{fig:example}(c). In Figures \ref{fig:example}(d) and \ref{fig:example}(e), MPE26 and MPE28 show morphological similarity between Pa$\alpha$ and H$\alpha$, despite no corresponding known \ion{H}{2} regions. No clear {\it WISE} counterparts of MPE26 and MPE28 appear in Figure \ref{fig:example}(f). Except for two Pa$\alpha$ point-like sources (MPP11 and MPP15) that have no available IPHAS data, all the other Pa$\alpha$ point-like sources have their IPHAS H$\alpha$ counterparts, as given in Table \ref{table:mpp}. In particular for MPP10, MPP12, and MPP18, each source was found to harbor not only a compact H$\alpha$ point source, but also a diffuse H$\alpha$ feature surrounding the point source in the IPHAS H$\alpha$ images. However, due to the lower resolution of MIPAPS, we could not resolve the individual diffuse H$\alpha$ features in the Pa$\alpha$ images.

We performed aperture photometry to extract Pa$\alpha$ total fluxes for 18 Pa$\alpha$ extended sources that show complete circular morphologies without overlapping with other bright sources. Except for two sources that were not fully covered by the IPHAS observation, we also carried out aperture photometry of H$\alpha$ total fluxes, and derived $E(\bv)$ values for the 16 Pa$\alpha$ extended sources. The methods of flux photometry and $E(\bv)$ estimation are the same as those described in Section \ref{subsec:wise_sources}, and the results are given in Table \ref{table:mpe}, together with 1$\sigma$ uncertainties. Although the precise photometry of point-like sources should be made by PSF-fitted photometry, we performed rough aperture photometry of Pa$\alpha$ fluxes for 16 Pa$\alpha$ point-like sources that do not largely overlap with stellar residual or other bright sources. To measure the total fluxes of the individual sources, we applied a sufficiently large aperture size (6 arcmin; $\sim$7 MIRIS pixels) to the point-like sources. Table \ref{table:mpp} gives the results, together with the 1$\sigma$ uncertainties.

\subsection{$E(\bv)$ map of Sh2-131} \label{subsec:bvmaps}

\begin{figure*}
\centering
\includegraphics[scale=0.6]{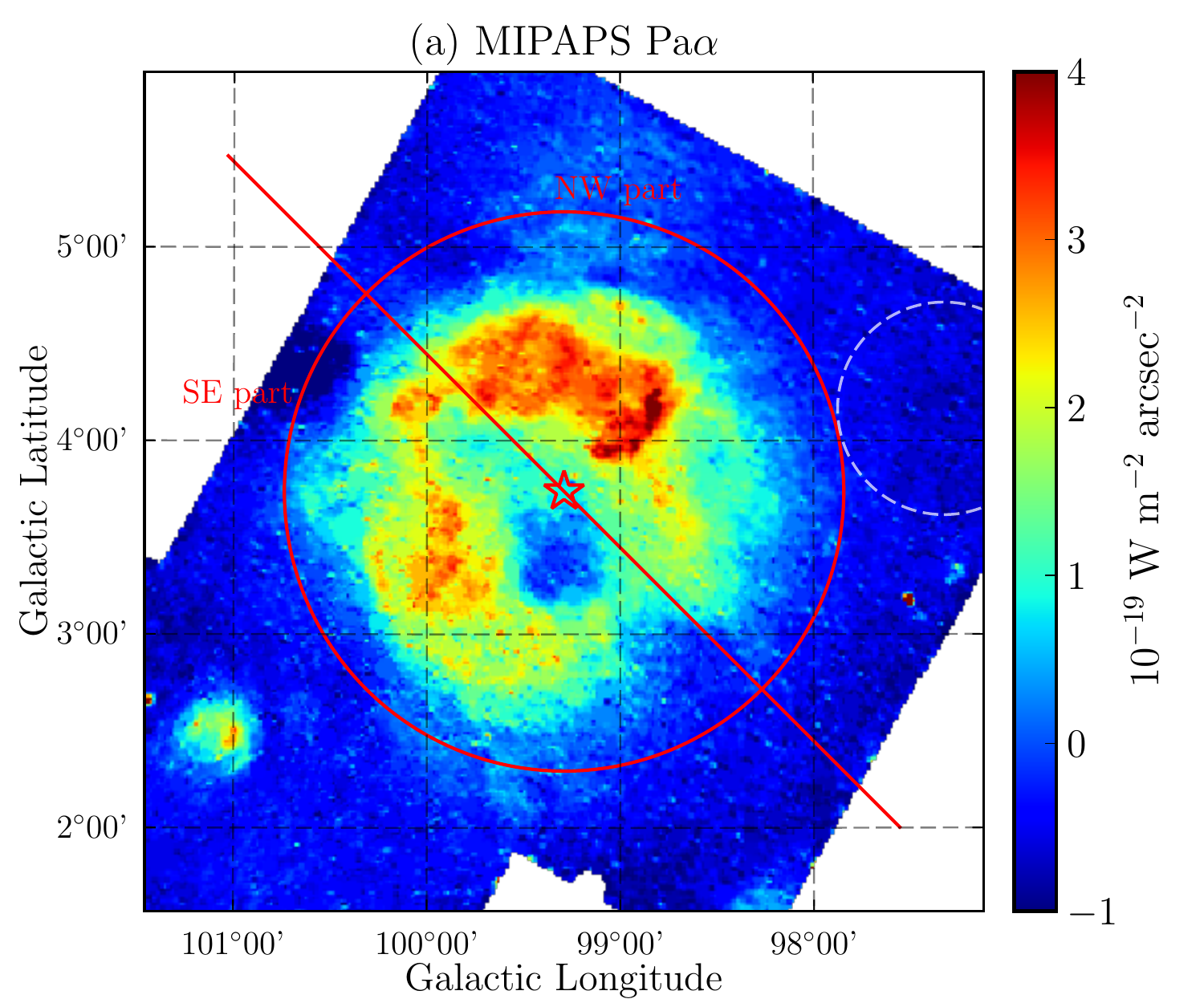}\includegraphics[scale=0.6]{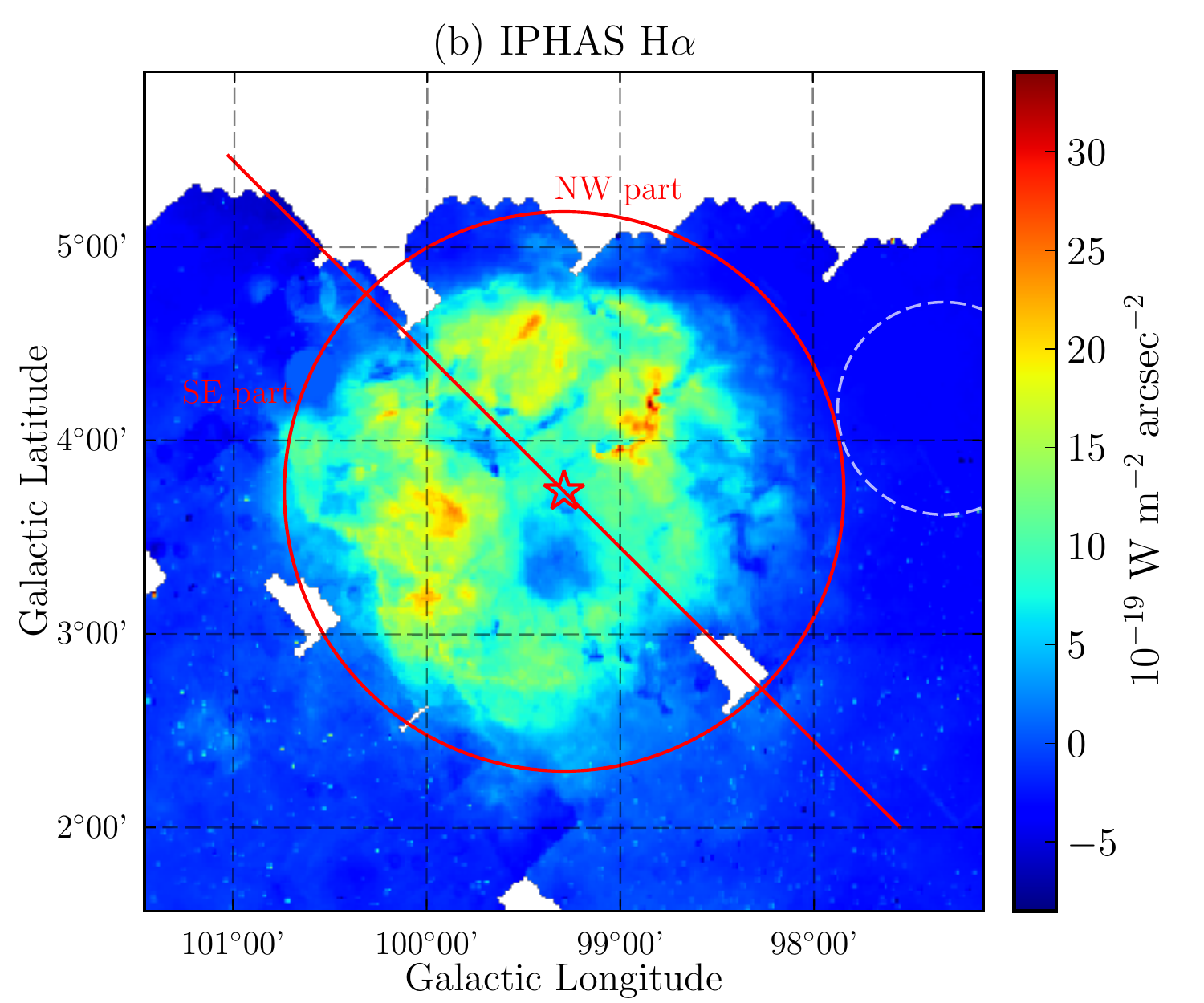}
\includegraphics[scale=0.6]{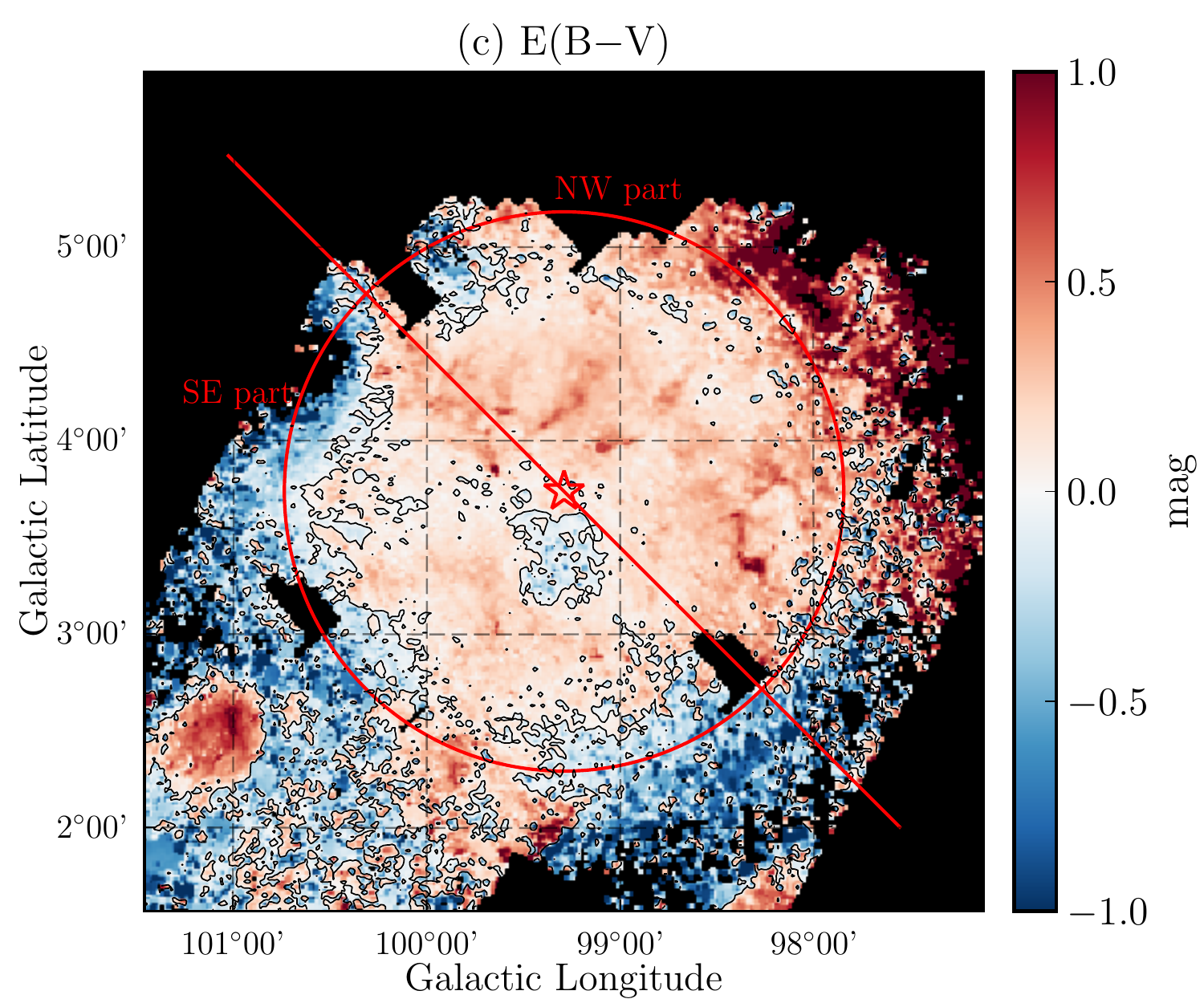}
\caption{(a) MIPAPS Pa$\alpha$, (b) IPHAS H$\alpha$, and (c) $E(\bv)$ images of WK03 (known as Sh2-131, at a distance of 1.00 $\pm$ 0.08 kpc by \citet{foster15}) in Galactic coordinates. The upper two images are close-up images extracted from Figures \ref{fig:paa} and \ref{fig:ha}, respectively. The IPHAS H$\alpha$ image was re-binned to match its pixel size and pixel positions with those of the MIPAPS Pa$\alpha$ image. The white dashed circles denote the area selected for background subtraction in each image. The lower image for $E(\bv)$ was derived from the two upper images (with the same pixel size of $\sim$52 arcsec), and the overlaid contours indicate a level of $E(\bv)$ = 0 mag. The red star symbols indicate the location of HD 206267, which is known to be a dominant ionizing star of WK03. The red solid circles centered on the star and the red diagonal lines divide the source region into two sub-regions: the northwest (NW) and the southeast (SE) parts. The radial profiles of $E(\bv)$ in Figure \ref{fig:wk03rp} were derived from these sub-regions. A compact feature shown in the lower left corner is WC14, which was classified in the {\it WISE} \ion{H}{2} region catalog as an \ion{H}{2} region candidate \citep{anderson14}.\label{fig:wk03map}}
\end{figure*}

WK03 (known as Sh2-131) is one of the very close \ion{H}{2} regions with a distance of 1.00 $\pm$ 0.08 kpc \citep{foster15} in the Cepheus region. As a result, we could identify the largest circular feature with a diameter of $\sim$2.\arcdeg5 in the Pa$\alpha$ and H$\alpha$ images (see Figures \ref{fig:paa} and \ref{fig:ha}), and obtain the lowest $E(\bv)$ of 0.2 mag (see Table \ref{table:wk}). In order to make an $E(\bv)$ map of this source, we extracted close-up images around WK03 from Figures \ref{fig:paa} and \ref{fig:ha}. We masked out stellar residuals around WK03, and adjusted the IPHAS H$\alpha$ image to match its pixel size and positions with those of the MIPAPS Pa$\alpha$ image. Figures \ref{fig:wk03map}(a) and \ref{fig:wk03map}(b) show the resulting Pa$\alpha$ and H$\alpha$ images of WK03, respectively. We estimated negative background values of -0.63 for Pa$\alpha$ and -3.68 for H$\alpha$ by taking the median values within the white dashed circles denoted in the figures, and subtracted the values from both images, respectively. Then, $E(\bv)$ at each pixel was derived by comparing the observed Pa$\alpha$ to H$\alpha$ ratio with the intrinsic value, as described in Section \ref{subsec:wise_sources}. Figure \ref{fig:wk03map}(c) shows the resulting $E(\bv)$ map. The $E(\bv)$ map of WK03 reveals higher extinctions in the northwest (NW) part of the \ion{H}{2} region, and lower extinctions in the southeast (SE) part. We also note that several filamentary features with the highest $E(\bv)$ values appear in the NW part. The contours corresponding to $E(\bv) = 0$ divide the regions with positive color excesses (red) from those with negative values (blue). It is noted that there are regions with negative $E(\bv)$ values at the center and the outer rim in the SE part. A compact object with high $E(\bv)$ values in the lower left corner of the figure is WC14, which was classified in the {\it WISE} \ion{H}{2} region catalog as an \ion{H}{2} region candidate \citep{anderson14}. The great distance to WC14 (4.87 $\pm$ 0.04 kpc by \citet{foster15}) may result in a very high $E(\bv)$ value and a small angular size, compared to those of WK03.

\section{Discussion} \label{sec:discussion}

\subsection{Visual Inspection} \label{subsec:visual}

We presented the maps of the Cepheus region measured at Pa$\alpha$ and H$\alpha$, and newly identified a total of 90 sources (54 ``Candidate'', 18 ``Group'', and 18 ``Radio Quiet'' sources) in the {\it WISE} \ion{H}{2} region catalog as true \ion{H}{2} regions, as summarized in Table \ref{table:vis}. Thus, they can now be classified as ``Known'' sources. Among those 90 sources, Pa$\alpha$ was detected from a total of 53 sources (39 ``Candidate'', 12 ``Group'', and 2 ``Radio Quiet'' sources), as given in the table, but the remaining 37 sources were only detected at H$\alpha$. As mentioned in Section \ref{sec:intro}, the less-attenuated Pa$\alpha$ flux becomes larger than the more-attenuated H$\alpha$ flux when $E(\bv)$ $>$1.12. Therefore, it was expected that Pa$\alpha$ could be detected from a lot of sources that are invisible at H$\alpha$ due to dust extinction. However, ignoring WK14, WG03, and WG05 which are located outside the IPHAS survey area, among the {\it WISE} \ion{H}{2} region sources invisible in the IPHAS H$\alpha$ images, only one source (WC55) was detected at Pa$\alpha$. In Tables \ref{table:wk}--\ref{table:wr}, only 23 sources have the spatially averaged $E(\bv)$ of $>$1.12, among the 62 sources with the estimated $E(\bv)$ values. Moreover, only 4 sources have $E(\bv)$ of $>$2.0. The Cepheus region analyzed in this study lies in the direction of the outer Galaxy ($\ell = 96\arcdeg.5$--$116\arcdeg.3$), which means that dust extinction is not as severe as that in the inner Galaxy. This is why IPHAS with higher spatial resolution and sensitivity has an advantage over MIPAPS for the detection of \ion{H}{2} regions in the Cepheus region. We note that MIPAPS was performed with only 8 cm aperture of the MIRIS telescope, compared to IPHAS with a 2.5 m telescope. However, the MIPAPS Pa$\alpha$ data are still believed to be useful in detecting \ion{H}{2} regions with high extinction of $E(\bv)$ $>$1.12, many of which are likely to reside in regions toward the inner Galaxy. Since more than half of the whole {\it WISE} \ion{H}{2} region candidates in the entire Galactic plane are in the inner Galactic region, we expect that the MIPAPS Pa$\alpha$ data would be more useful to identify them as definite \ion{H}{2} regions than optical H$\alpha$ data. In particular, for the sources with large angular sizes, the MIPAPS Pa$\alpha$ images can provide better information for high-extinction regions than H$\alpha$. For example, the MIPAPS Pa$\alpha$ image of WK26 seen in Figure \ref{fig:example}(d) shows an almost complete circular feature, but only bright parts in the north side are identified in the IPHAS H$\alpha$ image, as shown in Figure \ref{fig:example}(e).

The most up-to-date version 2.0 of the catalog is presented in the project homepage for the {\it WISE} \ion{H}{2} region catalog\footnote{\url{http://astro.phys.wvu.edu/wise/}}. We note that 20 ``Candidate'' and 2 ``Group'' sources, among the \ion{H}{2} region candidates in the Cepheus region, were updated to ``Known'' sources, which are indicated in Tables \ref{table:wc} and \ref{table:wg}. \citet{anderson15} and \citet{anderson18} detected hydrogen RRLs from them using the Green Bank telescope. All of the 22 sources are included in the 90 \ion{H}{2} regions newly confirmed from this study.

\citet{anderson14} claimed that the {\it WISE} \ion{H}{2} region catalog is the most complete catalog of Galactic \ion{H}{2} regions, based on the expected MIR brightness of \ion{H}{2} regions, and the sensitivity of the {\it WISE} data. However, they classified only well-defined circular sources as distinct \ion{H}{2} regions, and excluded irregular ionized sources with incomplete structures, or complexes composed of multiple objects. Figures \ref{fig:paa} and \ref{fig:ha} show that there are lots of large-extent diffuse features that are visible at both Pa$\alpha$ and H$\alpha$. Although most of them are not included in the {\it WISE} \ion{H}{2} region catalog, morphological similarity between Pa$\alpha$ and H$\alpha$ indicates that they are real features tracing ionized hydrogen gas. They can be a part of nearby diffuse \ion{H}{2} regions, even though their incomplete morphological characteristics. As mentioned in Section \ref{subsec:mipaps_sources}, 16 Pa$\alpha$ extended sources identified in this study were found to have known \ion{H}{2} region entries in the SIMBAD database, although they are not included in the {\it WISE} \ion{H}{2} region catalog. The other 13 Pa$\alpha$ extended sources with no corresponding known \ion{H}{2} regions may also be related to \ion{H}{2} regions or ionized sources. MPE01, MPE19, and MPE28 may just be the features related to adjacent bright sources, one of which (MPE28) is shown in Figures \ref{fig:example}(d)--\ref{fig:example}(f). However, they might be separate \ion{H}{2} regions, as in the case of MPE15 in Figures \ref{fig:example}(a)--\ref{fig:example}(c). In Figures \ref{fig:example}(d)--\ref{fig:example}(f), MPE26 with a plausible filamentary feature, can be a part of an \ion{H}{2} region, or a feature related to an unknown shock front. We note that the remaining 9 Pa$\alpha$ extended sources show isolated circular features at Pa$\alpha$ and H$\alpha$, and thus they can be newly identified \ion{H}{2} regions.

As listed in Table \ref{table:mpp}, we detected Pa$\alpha$ emissions from 3 planetary nebulae and 15 emission-line stars, including 6 Wolf-Rayet and 2 Herbig Ae/Be stars in the Cepheus region. All of them, except for two sources located outside the survey area, have their IPHAS counterparts. In particular, diffuse extended H$\alpha$ features together with H$\alpha$ point sources were found for 3 emission-line stars (MPP10, MPP12, and MPP18) in the IPHAS H$\alpha$ image, as mentioned in Section \ref{subsec:mipaps_sources}. The central source (MWC 1080) related to MPP18 is classified as a Herbig Ae/Be star with B0 spectral type in SIMBAD, and thus the surrounding extended H$\alpha$ feature is likely an initial \ion{H}{2} region formed by MWC 1080. The central source (V669 Cep) related to MPP10 was also reported as a Herbig Ae/Be star by \citet{chen16}, although the spectral type of V669 Cep was presented as B5 (too late to form an \ion{H}{2} region). If V669 Cep has a slightly earlier spectral type, the extended H$\alpha$ feature around V669 Cep can also be considered to be an initial \ion{H}{2} region formed by V669 Cep. Otherwise, it may be an ionized hydrogen feature related to some violent activities of the Herbig Ae/Be star, such as outflows and jets emitted from the Herbig star. In fact, MPP16 (known as AS 505), which is classified as a Herbig Ae/Be star with B5 spectral type in SIMBAD, is surrounded by a {\it WISE} \ion{H}{2} region candidate WC56, from which both Pa$\alpha$ and H$\alpha$ were detected.

\begin{figure*}
\centering
\includegraphics[scale=0.7]{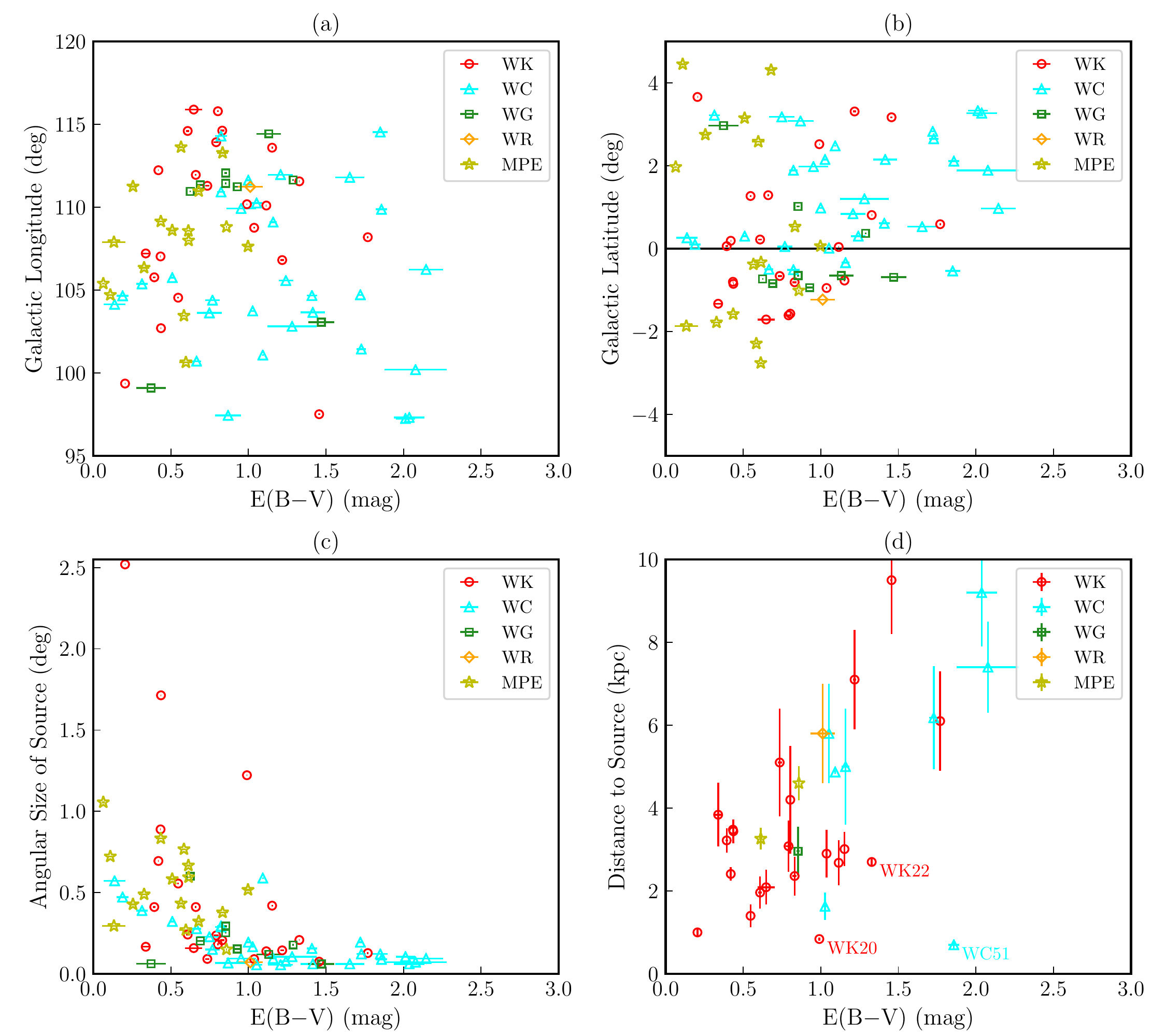}
\caption{$E(\bv)$ vs. (a) Galactic longitude, (b) Galactic latitude, (c) angular size of source, and (d) distance to source for 78 sources, of which the $E(\bv)$ values are available in Tables \ref{table:wk}--\ref{table:wr} and \ref{table:mpe}. ``WK'', ``WC'', ``WG'', and ``WR'' represent {\it WISE} ``Known'', ``Candidate'', ``Group'', and ``Radio Quiet'' sources, respectively. ``MPE'' represents MIPAPS Pa$\alpha$ extended sources. The $E(\bv)$ values (with 1$\sigma$ error bars) are from the ``$E(\bv)$'' columns of the tables. In (d), only 33 sources with known distances (including their uncertainties) in Table \ref{table:dis} are shown together with vertical error bars for the distances. The symbols without visible error bars denote the sources with uncertainties that are smaller than the symbol sizes.\label{fig:cecorr}}
\end{figure*}

\subsection{Photometric Results} \label{subsec:photometric}

From the photometries of the MIPAPS Pa$\alpha$ and IPHAS H$\alpha$ data, we obtained the Pa$\alpha$-H$\alpha$ $E(\bv)$ color excesses for 62 {\it WISE} \ion{H}{2} region sources (22 ``Known'', 30 ``Candidate'', 9 ``Group'', and 1 ``Radio Quiet'' sources) and 16 MIPAPS Pa$\alpha$ extended sources, as given in Tables \ref{table:wk}--\ref{table:wr} and \ref{table:mpe}. We compare the $E(\bv)$ values with Galactic coordinates, angular sizes of the sources, and distances to the sources in Figure \ref{fig:cecorr}. In Figure \ref{fig:cecorr}(d), only 33 sources with known distances are plotted. We collected kinematic, spectrophotometric, or parallax distance values for the 33 sources from \citet{anderson14}, \citet{foster15}, or \citet{moscadelli09}, which are shown in Table \ref{table:dis}. Regarding the sources with multiple distance values, we adopted the value with the highest distance-to-uncertainty ratios (those without parentheses in Table \ref{table:dis}), and used them in Figure \ref{fig:cecorr}(d). No significant correlation with Galactic coordinates is found, whereas the dependency of $E(\bv)$ on angular size and distance is clearly found. The size of the Cepheus region dealt with in this study is less than 20 degrees, which might be too small to see the dependency of $E(\bv)$ on Galactic coordinates. In principle, the negative and positive correlation with angular size and distance, respectively, can be explained by the fact that the more distant sources are in general smaller in angular size and more attenuated by interstellar dust. However, not only the more distant \ion{H}{2} regions, but also young ultracompact \ion{H}{2} regions still embedded in dense molecular clouds could have smaller angular sizes. The embedded \ion{H}{2} regions will suffer from higher internal extinction by their own local clouds, even when they are attenuated by only a small amount of interstellar dust between them and us. Therefore, the sources located at the lower right part (high $E(\bv)$ and small distance) in Figure \ref{fig:cecorr}(d) are likely such young ultracompact \ion{H}{2} regions. For example, WC51 with an $E(\bv)$ of 1.86 $\pm$ 0.04 mag and a distance of 0.7 $\pm$ 0.1 kpc would be the case. The source has not been reported in previous studies for the H$\alpha$ detection, probably due to high extinction, and thus was classified as a ``Candidate'' source in the {\it WISE} \ion{H}{2} region catalog. As denoted in the figure, WK22 shows relatively strong extinction as well. It is known to be an \ion{H}{2} region, Sh2-158 (also known as NGC 7538), which was also reported to be associated with a giant molecular cloud, G111.50+0.75 (cloud G) by \citet{ungerechts00}. \citet{moscadelli09} reported the detections of 12 GHz methanol masers toward WK22 and WC51 (NGC 7538 and Cep A in their paper, respectively). Since methanol masers are well-known indicators of the early phases of massive star formation \citep{urquhart13}, the detections of methanol masers for WK22 and WC51 support the suggestion that these sources located at the lower right part of Figure \ref{fig:cecorr}(d) are likely young ultracompact \ion{H}{2} regions embedded in dense molecular clouds.

\begin{figure}
\centering
\includegraphics[scale=0.7]{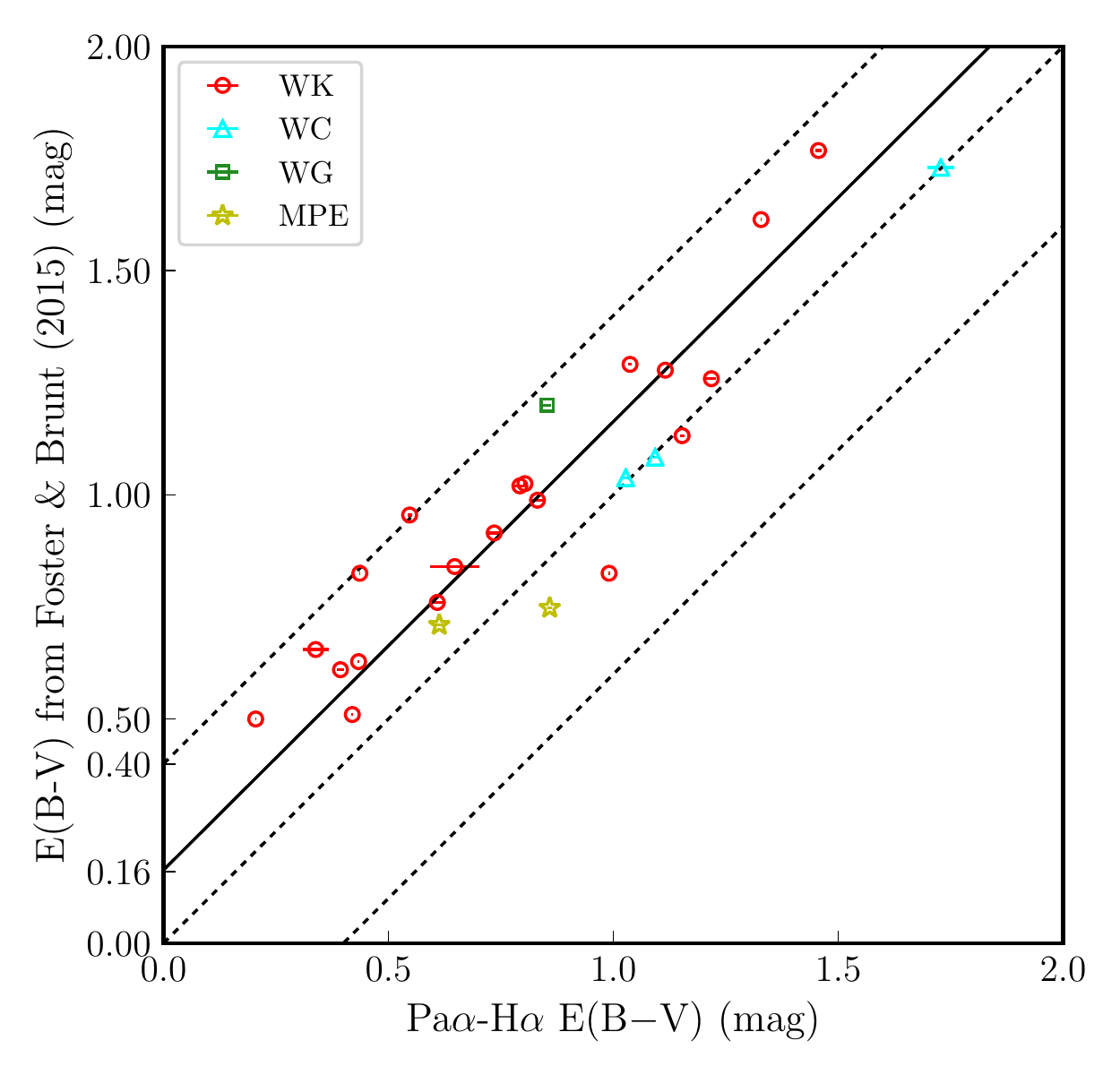}
\caption{Comparison of our Pa$\alpha$-H$\alpha$ $E(\bv)$ with the $E(\bv)$ values presented by \citet{foster15} for 26 sources, of which the counterparts are available in \citet{foster15}, as listed in Table \ref{table:dis}. ``WK'', ``WC'', and ``WG'' represent {\it WISE} ``Known'', ``Candidate'', and ``Group'' sources, respectively. ``MPE'' represents MIPAPS Pa$\alpha$ extended sources. The Pa$\alpha$-H$\alpha$ $E(\bv)$ values (with 1$\sigma$ error bars) are from the ``$E(\bv)$'' columns in Tables \ref{table:wk}--\ref{table:wg} and \ref{table:mpe}. The symbols without visible error bars denote the sources with the Pa$\alpha$-H$\alpha$ $E(\bv)$ uncertainties that are smaller than the symbol sizes. The $E(\bv)$ values of \citet{foster15} (without uncertainties) were taken from their Table 2, as shown in Table \ref{table:dis}. The diagonal dotted lines show agreement between the two kinds of $E(\bv)$ within $\sim$0.4 mag. On the other hand, the solid line represents the best-fit line with a systematic offset of 0.16 $\pm$ 0.03 mag.\label{fig:cecomp}}
\end{figure}

\citet{foster15} calculated the $E(\bv)$ values for 103 \ion{H}{2} regions, using photometric and spectroscopic information of the stars associated with the \ion{H}{2} regions. Among the sources dealt with by \citet{foster15}, we found a total of 26 sources corresponding to our sources with the Pa$\alpha$-H$\alpha$ $E(\bv)$ values in Tables \ref{table:wk}--\ref{table:wr} and \ref{table:mpe}, which are shown in Table \ref{table:dis}. By checking positions and angular sizes of the sources with the same names in both \citet{anderson14} and \citet{foster15}, we found two \ion{H}{2} regions WK28 and WK29 (S166 and S165 in \citet{anderson14}, respectively) to be mistakenly interchanged in the {\it WISE} \ion{H}{2} region catalog, and thus we related WK28 and WK29 to Sh2-165 and Sh2-166, respectively. Sh2-161 in \citet{foster15} seems to correspond more likely to WG14, rather than to WK24. Since WK21 is an ultracompact \ion{H}{2} region existing within the larger \ion{H}{2} region, Sh2-157, we adopted only the $E(\bv)$ value for the star, ALS 19704, in Table 1 of \citet{foster15}, which positionally matches with the center of WK21. We also used only LS I +6050 for WK30, because the other ionizing stars of Sh2-168 presented by \citet{foster15} are located outside our angular boundary of WK30. Additionally, three ``Candidate'' {\it WISE} sources (WC14, WC15, and WC27) and two MIPAPS Pa$\alpha$ extended sources (MPE13 and MPE15) were found to have their counterparts in \citet{foster15}, as given in Table \ref{table:dis}.

For the above 26 sources, we compared our Pa$\alpha$-H$\alpha$ $E(\bv)$ with the $E(\bv)$ values presented by \citet{foster15} in Figure \ref{fig:cecomp}, which shows good agreement within $\sim$0.4 mag, as denoted by diagonal dotted lines in the figure. However, interestingly, it is found that the Pa$\alpha$-H$\alpha$ $E(\bv)$ values are systematically lower than those by \citet{foster15}. By the least-square linear fitting, we estimated the offset of the Pa$\alpha$-H$\alpha$ $E(\bv)$ from those by \citet{foster15}, and then obtained the value of 0.16 $\pm$ 0.03 mag, as indicated by a diagonal solid line in Figure \ref{fig:cecomp}. We note that the Pa$\alpha$-H$\alpha$ $E(\bv)$ values were directly derived from the photometry of extended emissions from ionized hydrogen gas in the \ion{H}{2} regions. On the other hand, the $E(\bv)$ values by \citet{foster15} were calculated from the photometry of point stars that are believed to be associated with the \ion{H}{2} regions. The deviation of the Pa$\alpha$-H$\alpha$ $E(\bv)$ from that by \citet{foster15} may originate from the difference of geometries of these target sources. The distance to each location of an extended \ion{H}{2} region would be different from those to the stars chosen to calculate $E(\bv)$, which may explain the random deviation of the $E(\bv)$ values. However, the geometrical difference cannot be a main cause of the systematic difference. Instead, the more plausible cause of the systematic underestimation of the Pa$\alpha$-H$\alpha$ $E(\bv)$ values is the effect of dust scattering. Figures 1 and 5 of \citet{seon12} show that a portion of light from an extended source can be scattered by dust into sightlines within the angular size of the source, and then this scattered light will be observed together with the direct light from the source. Since the H$\alpha$ line is at a shorter wavelength than the Pa$\alpha$ line, the dust-scattering effect is more important at H$\alpha$ than at Pa$\alpha$, and thus the scattering component is stronger at H$\alpha$ than at Pa$\alpha$. Therefore, the Pa$\alpha$-H$\alpha$ $E(\bv)$, which is defined by the ratio between the Pa$\alpha$ and H$\alpha$ total fluxes from an extended source (i.e., \ion{H}{2} region), will underestimate its true value by as much as is contributed by the difference in dust scattering. On the other hand, for a point source, the scattered light is completely missed from the sightline of the source, and thus no scattered light is detected. Therefore, the $E(\bv)$ values estimated from point stars are free from the dust-scattering effect. In the case of Figure \ref{fig:cecomp}, dust scattering seems to contribute on average 0.16 $\pm$ 0.03 mag to the Pa$\alpha$-H$\alpha$ $E(\bv)$ values. However, this value can vary somewhat, depending on the assumed temperature of ionized hydrogen gas. If a temperature of 5 $\times$ 10$^{3}$ K or 2 $\times$ 10$^{4}$ K (instead of 10$^{4}$ K) in Table 14.2 of \citet{draine11} is applied, the offset is calculated to be 0.23 $\pm$ 0.03 mag or 0.10 $\pm$ 0.03 mag, respectively.

Using two three-dimensional (3D) extinction data, given by \citet{sale14} and \citet{green15}, we also attempted to derive the $E(\bv)$ values for the sources, of which the known distances are available in Table \ref{table:dis}. The 3D extinction data of \citet{sale14} were obtained with an angular resolution of $\sim$600 arcsec and a distance resolution of 100 pc, by using the IPHAS photometry of stars. The 3D extinction data of \citet{green15} were derived from Pan-STARRS 1 optical and 2MASS near-infrared photometry of stars with an angular resolution of 204--822 arcsec and a distance resolution of 0.5 distance modulus ($\sim$12.6 pc). Since some sources with small angular sizes have no reliable 3D extinction data within their areas, the $E(\bv)$ values for only 20 and 23 sources were obtained from the 3D data of \citet{sale14} and \citet{green15}, respectively. In comparison with the Pa$\alpha$-H$\alpha$ $E(\bv)$ values, the discrepancy was within $\sim$0.6 mag. However, there was no clear systematic offset between the Pa$\alpha$-H$\alpha$ $E(\bv)$ and those derived from the 3D data. In fact, the $E(\bv)$ values from the 3D data have large uncertainties, due to inaccurate values of distances to the sources, which were used to calculate $E(\bv)$ from the 3D data. The sources with small angular sizes could be affected by the limited angular resolutions of the 3D data, as well.

\begin{figure*}
\centering
\includegraphics[scale=0.7]{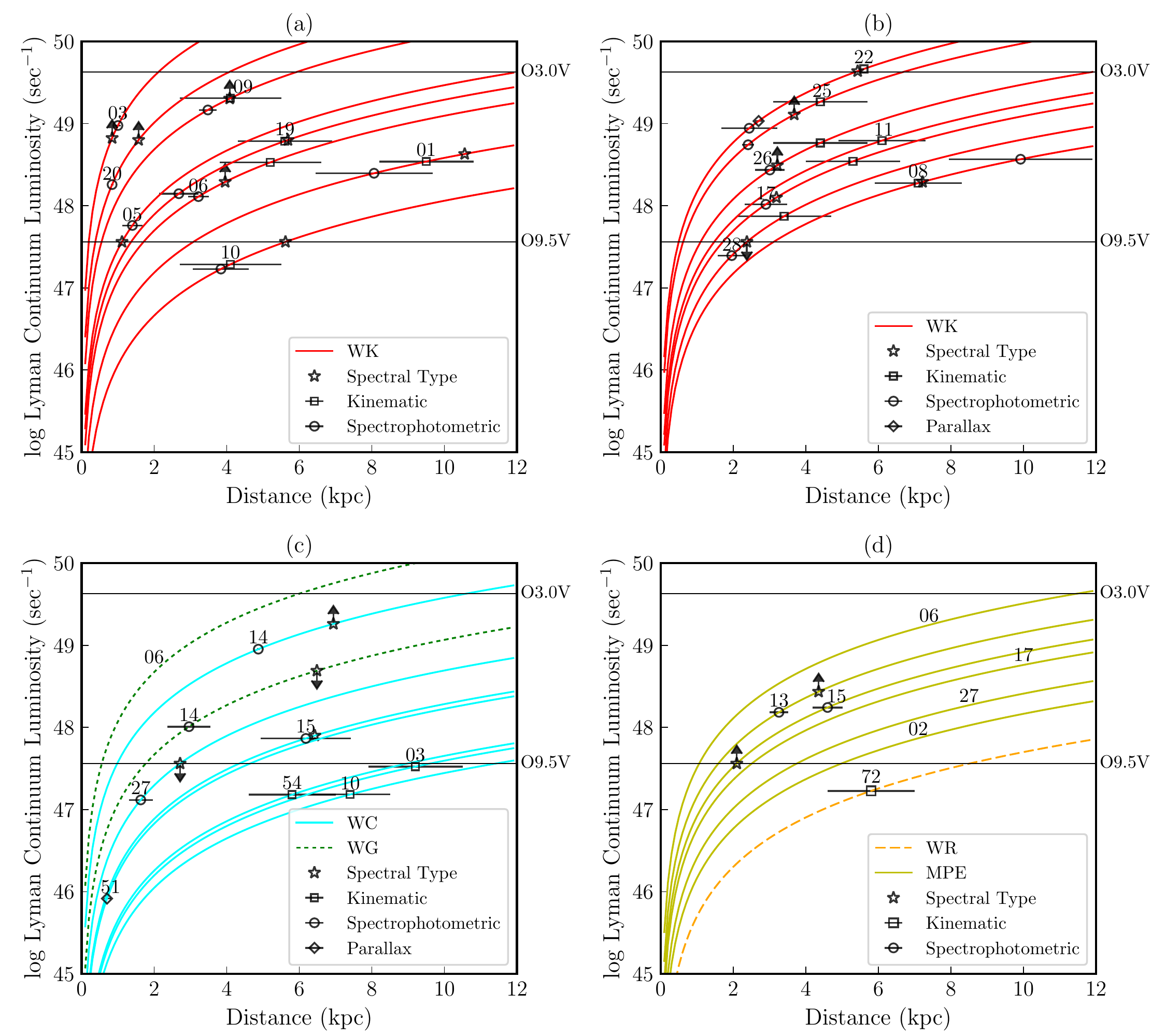}
\caption{Total Lyman continuum luminosity as a function of distance for 31 representative sources. ``WK'', ``WC'', ``WG'', and ``WR'' represent {\it WISE} ``Known'', ``Candidate'', ``Group'', and ``Radio Quiet'' sources, respectively. ``MPE'' represents MIPAPS Pa$\alpha$ extended sources. The curve for each source (the numbers on the curves denote the source ID numbers given in Tables \ref{table:wk}--\ref{table:wr} and \ref{table:mpe}) is calculated by using its Pa$\alpha$ total flux and $E(\bv)$ given in Tables \ref{table:wk}--\ref{table:wr} and \ref{table:mpe}. Utilizing Table 1 of \citet{martins05}, the positions corresponding to Lyman continuum luminosities for O3.0V- and O9.5V-type stars are indicated by two horizontal lines. Star symbols on the curves indicate the positions corresponding to Lyman continuum luminosities calculated using Tables 1--3 of \citet{martins05} and the spectral types of the ionizing stars given in Table \ref{table:dis}. The stars with the upper (or lower) limit symbols represent the sources that have only one B-type ionizing star (or additional B-type ionizing stars, as well as O-type ionizing stars). Squares, circles, and diamonds (with error bars) on the curves denote the positions corresponding to the kinematic, spectrophotometric, and parallax distances given in Table \ref{table:dis}, respectively. \label{fig:lyc}}
\end{figure*}

To estimate the intrinsic total flux at Pa$\alpha$ or H$\alpha$ for each \ion{H}{2} region, its spatially-averaged $E(\bv)$ value can be used for the observed total flux. Using the Pa$\alpha$-H$\alpha$ $E(\bv)$ value and the Pa$\alpha$ total flux observed by MIRIS, we obtained the reddening-corrected Pa$\alpha$ total flux for each \ion{H}{2} region. Assuming the relative luminosity fractions between the recombination lines, as shown in Table 14.2 of \citet{draine11} for the case B condition, the total luminosity of the Lyman continuum emitted from its ionizing star(s) can be derived from the intrinsic Pa$\alpha$ total flux (once a distance to an \ion{H}{2} region is known). Then, the spectral type(s) of the star(s) can also be inferred by comparing the derived luminosity with the theoretically predicted value \citep{martins05}. Figure \ref{fig:lyc} shows the total Lyman continuum luminosity as a function of distance for 31 representative sources, of which the Pa$\alpha$-H$\alpha$ $E(\bv)$ values are available. Squares, circles, and a diamond (with error bars for distances) on the curves indicate the positions corresponding to kinematic, spectrophotometric, and parallax distances given in Table \ref{table:dis}, respectively. Since the kinematic distance for WK22 has no uncertainty value, it is denoted in Figure \ref{fig:lyc}(b) by a square without its error bar. Star symbols on the curves indicate the positions corresponding to Lyman continuum luminosities calculated using Tables 1--3 of \citet{martins05} and the spectral types of the ionizing stars given in Table \ref{table:dis}. Because \citet{martins05} presented Lyman continuum luminosities for only O3- to O9.5-types, we considered only O-type ionizing stars to calculate Lyman continuum luminosities. Instead, we provide the upper or lower limits of the Lyman continuum luminosities for the \ion{H}{2} regions with B-type ionizing stars. In addition, we assumed V for the sources with luminosity classes of IV and (V) in Table \ref{table:dis}, because \citet{martins05} presented Lyman continuum luminosities for only luminosity classes of I, III, and V. The stars with the upper limit symbols represent the sources that individually have only one B-type ionizing star, and the upper limits are the values for the latest O-type (O9.5) star with the corresponding luminosity class in \citet{martins05}. On the other hand, the stars with the lower limit symbols represent the sources that have additional B-type ionizing stars, as well as O-type ionizing stars, and the lower limits are the values obtained from only the O-type ionizing stars.

The total Lyman continuum luminosity calculated for each source can be used to constrain either the distance to the source or the spectral type(s) of its ionizing star(s), if we know the other one. For example, WK22 (known as Sh2-158) is known to have a parallax distance of 2.7 $\pm$ 0.1 kpc \citep{moscadelli09,anderson14}, a spectrophotometric distance of 2.44 $\pm$ 0.77 kpc \citep{foster15}, and two ionizing stars with spectral types of O3V and O9V \citep{russeil07,foster15}, as shown in Table \ref{table:dis}. In Figure \ref{fig:lyc}(b), the positions corresponding to the parallax distance of 2.7 kpc and the spectrophotometric distance of 2.44 kpc are indicated on the curve of WK22 by a diamond and a circle, respectively. On the other hand, a star symbol is plotted at the position corresponding to the Lyman continuum luminosity emitted from two stars with spectral types of O3V and O9V. If the distance of 2.7 kpc or 2.44 kpc is correct, then the spectral types of its ionizing stars should be later than O3V and O9V. In another study, \citet{lynds86} considered only a central O7V-type star to be the main ionizing star of WK22. On the other hand, if the two stars presented by \citet{russeil07} and \citet{foster15} are indeed O3V- and O9V-type ionizing stars for WK22, then the distance to WK22 is likely to be $\sim$5.4 kpc, as shown in the figure. This is similar to the kinematic distance of WK22 (5.6 kpc, as denoted by a square in the figure), although \citet{moscadelli09} concluded that the parallax distance of 2.7 kpc obtained from measuring methanol masers is the actual distance to WK22. In the case of WK20, the total luminosity of its ten ionizing stars suggests a much larger distance ($\ga$1.6 kpc) than the spectrophotometric distance of 0.84 $\pm$ 0.04 kpc, as shown in Figure \ref{fig:lyc}(a). This implies that the relatively low distance for WK20 compared to $E(\bv)$ in Figure \ref{fig:cecorr}(d) can be simply due to adopting the underestimated distance value of 0.84 kpc. We note that the total Lyman continuum luminosities estimated from the kinematic distances to WK08, WK09, WK19, and WK22 are coincident with those estimated from the spectral types of their ionizing stars, as can be seen in Figure \ref{fig:lyc}. On the other hand, the total Lyman continuum luminosities estimated from the spectrophotometric distances to WK03, WK17, WK26, WK28, and WC15 are compatible with those estimated from the spectral types of their ionizing stars. We also note that the total Lyman continuum luminosities in Figure \ref{fig:lyc} could be somewhat underestimated, because the Pa$\alpha$-H$\alpha$ $E(\bv)$ values derived from extended components were found to be systematically underestimated by the dust-scattering effect, as discussed above.

The Pa$\alpha$-H$\alpha$ $E(\bv)$ values for the \ion{H}{2} regions in Cepheus show a general trend that the more distant sources have higher dust extinction. However, the Pa$\alpha$-H$\alpha$ $E(\bv)$ values estimated for the ``extended'' regions are found to be systematically lower than the $E(\bv)$ estimated using ``point'' stars by 0.10--0.23 mag. The discrepancy can be explained by the dust scattering effect which  enhances the H$\alpha$ flux from the extended sources. We note that the H$\alpha$ enhancement due to dust scattering affects only the Pa$\alpha$-H$\alpha$ $E(\bv)$ obtained for ``extended'' sources, but not the  $E(\bv)$ estimated for ``point'' sources. The discrepancy of 0.10--0.23 mag is found to be consistent with dust radiative transfer models (Seon et al. 2018, in preparation). Despite the effect of dust scattering, Figure \ref{fig:lyc} shows that the total Lyman continuum luminosities estimated for 9 \ion{H}{2} regions are simultaneously consistent with their known kinematic or spectrophotometric distances and the spectral types of their known ionizing stars. We will extend this photometric analysis to other regions in the Galactic plane, and compare the inner Galactic regions with higher extinctions and the outer Galactic regions with lower extinctions in subsequent papers.

\subsection{Morphological Results} \label{subsec:morphological}

Using the MIPAPS Pa$\alpha$ and IPHAS H$\alpha$ images, we made a Pa$\alpha$-H$\alpha$ $E(\bv)$ map for the whole region of WK03. Since WK03 is a large, nearby \ion{H}{2} region, we can examine the morphology in detail, even with the moderate spatial resolution of MIPAPS. \citet{foster15} noted that the O6.5V-type star HD 206267 is the dominant ionizing star for this \ion{H}{2} region. The position of the star is denoted by the star symbol in Figures \ref{fig:wk03map}(a)--\ref{fig:wk03map}(c). In the NW part of WK03, the H$\alpha$ image (Figure \ref{fig:wk03map}(b)) shows many filamentary dark features seen against bright environments, whereas these dark features are invisible in the Pa$\alpha$ image (Figure \ref{fig:wk03map}(a)). This results in filamentary features with high extinctions in the $E(\bv)$ image (Figure \ref{fig:wk03map}(c)), which reveal a morphology of dust clouds located in front of WK03. In the SE part with relatively low extinctions, a central void located just south of HD 206267 appears more clearly in the Pa$\alpha$ image than in the H$\alpha$ image, which results in negative $E(\bv)$ at that position. The contours of $E(\bv)$ = 0 mag in Figure \ref{fig:wk03map}(c) indicate negative $E(\bv)$ along the outer rim of the SE part, as well. The negative extinction means that there are additional H$\alpha$ emissions over those predicted from the intrinsic H$\alpha$-to-Pa$\alpha$ ratio. The additional H$\alpha$ component can be explained by the dust-scattering effect. \citet{seon12} demonstrated that the observed morphology of diffuse H$\alpha$ emission outside of the bright \ion{H}{2} regions accords well with the dust-scattering halo surrounding the \ion{H}{2} regions. They concluded that the dust-scattering effect is indeed important near the \ion{H}{2} regions.

\begin{figure}
\centering
\includegraphics[scale=0.68]{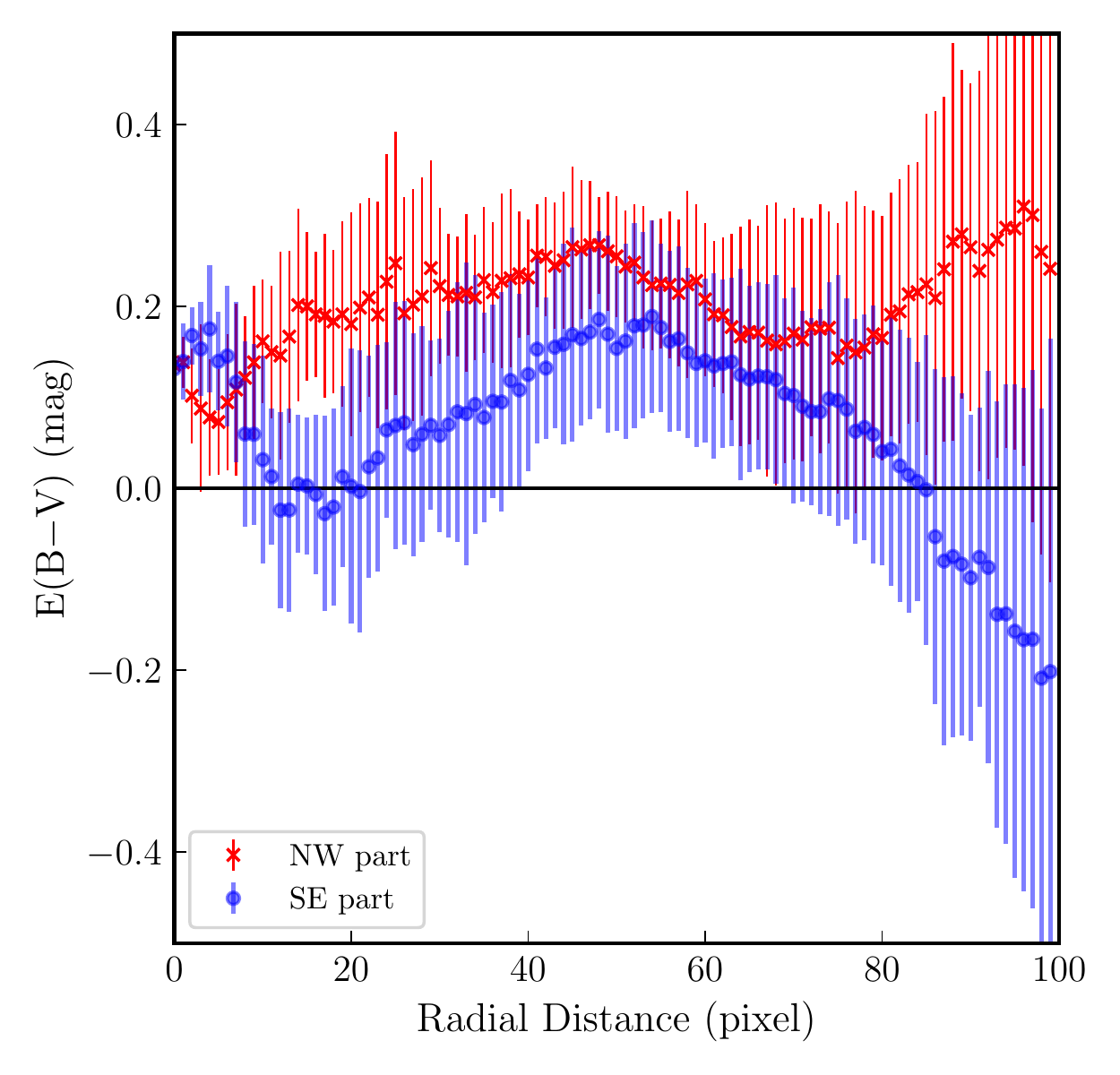}
\caption{Radial profiles of $E(\bv)$ for the northwest (NW) part (red x symbols with thin error bars), and the southeast (SE) part (blue circles with thick error bars) of WK03 based on Figure \ref{fig:wk03map}(c). The representative $E(\bv)$ for each radial distance was calculated as the median value for the pixels located at the radial distance within the solid circle in Figure \ref{fig:wk03map}(c).\label{fig:wk03rp}}
\end{figure}

Since the Pa$\alpha$ line (with a longer wavelength) is relatively free from dust scattering, the effect of dust scattering makes H$\alpha$ flux stronger than Pa$\alpha$ flux at the edge and outside of an \ion{H}{2} region. In order to verify this trend, we plotted the radial profiles of $E(\bv)$ for the WK03 region. The region, enclosed by the red solid circle centered on HD 206267, was divided into two sub regions (the NW part and the SE part) by the diagonal line in Figures \ref{fig:wk03map}(a)--\ref{fig:wk03map}(c). We calculated the representative $E(\bv)$ at each radial distance as the median value for the pixels located at the radial distance within the solid circle. Figure \ref{fig:wk03rp} shows the results. In contrast to the NW part (denoted by red x symbols with thin error bars), the radial profile of $E(\bv)$ in the SE part (denoted by blue circles with thick error bars) is rapidly decreasing near the edge of the source, which agrees with the behavior expected by dust scattering. Since this nearby \ion{H}{2} region basically has quite low interstellar extinction and its SE part has no foreground dust clouds, $E(\bv)$ values measured in the SE part can drop to negative values, which enables us to prove the dust-scattered H$\alpha$ components. Comparing the Pa$\alpha$ image with the H$\alpha$ image in Figures \ref{fig:wk03map}(a) and \ref{fig:wk03map}(b), we can know that diffuse and faint H$\alpha$ emissions are apparent even in regions invisible at Pa$\alpha$ (the central void and the outer rim of the SE part). These diffuse H$\alpha$ emissions could be light scattered by dust grains, which reside around and in front of the \ion{H}{2} region. Eventually, the MIPAPS Pa$\alpha$ morphology, which is much less affected by extinction and scattering, would be a better indicator of ionized hydrogen gas in this \ion{H}{2} region.

\section{Summary} \label{sec:summary}

Using the two narrow-band filters of MIRIS, the first Pa$\alpha$ imaging survey for the whole Galactic plane has been completed. To appreciate the data quality and scientific potential of the MIPAPS data, we analyzed the Pa$\alpha$ data for the Galactic longitude range of $\ell = 96\arcdeg.5$--$116\arcdeg.3$ around Cepheus, and together with the IPHAS H$\alpha$ data. We visually examined the Pa$\alpha$ and H$\alpha$ images for the 31 ``Known'', 71 ``Candidate'', 18 ``Group'', and 92 ``Radio Quiet'' sources in the {\it WISE} \ion{H}{2} region catalog. The Pa$\alpha$ and/or H$\alpha$ recombination lines were newly detected from a total of 90 sources (54 ``Candidate'', 18 ``Group'', and 18 ``Radio Quiet'' sources) out of 181 \ion{H}{2} region candidates. These sources can now be definitely identified as true \ion{H}{2} regions, or as ``Known'' sources according to the terminology of the {\it WISE} \ion{H}{2} region catalog. In particular, the Pa$\alpha$ emissions were detected from a total of 53 sources (39 ``Candidate'', 12 ``Group'', and 2 ``Radio Quiet'' sources). The Pa$\alpha$ emission line will have the advantage of being less attenuated compared to the H$\alpha$ line, in particular in regions with higher extinction, such as in the inner Galaxy. We also newly detected 29 extended and 18 point-like sources at Pa$\alpha$, which are not included in the {\it WISE} \ion{H}{2} region catalog. Their counterparts at H$\alpha$ were also found for all of the 29 extended sources and 16 point-like sources. In SIMBAD, it was found that 16 Pa$\alpha$ extended sources are associated with known \ion{H}{2} regions, and 18 Pa$\alpha$ point-like sources are due to 3 planetary nebulae and 15 emission-line stars, including 6 Wolf-Rayet and 2 Herbig Ae/Be stars. Out of the other 13 Pa$\alpha$ extended sources with no known counterparts, 9 Pa$\alpha$ extended sources are considered to be newly identified \ion{H}{2} regions, based on their isolated circular morphologies at Pa$\alpha$ and H$\alpha$.

By using the results of the aperture photometries of the MIPAPS Pa$\alpha$ and IPHAS H$\alpha$ total fluxes, we calculated the Pa$\alpha$-H$\alpha$ $E(\bv)$ color excesses for 62 {\it WISE} \ion{H}{2} region sources (22 ``Known'', 30 ``Candidate'', 9 ``Group'', and 1 ``Radio Quiet'' sources) and 16 MIPAPS Pa$\alpha$ sources. We found that the Pa$\alpha$-H$\alpha$ $E(\bv)$ values are basically in proportion to the distances to the sources, which is consistent with the fact that the more distant sources are commonly more attenuated by a larger amount of interstellar dust. Two sources with relatively high $E(\bv)$ compared to distance, could be young ultracompact \ion{H}{2} regions still embedded in dense molecular clouds. Also, we compared the Pa$\alpha$-H$\alpha$ $E(\bv)$ values of 26 \ion{H}{2} regions with those presented by \citet{foster15}. We note that the Pa$\alpha$-H$\alpha$ $E(\bv)$ values were derived from extended emissions of Pa$\alpha$ and H$\alpha$ originating from the \ion{H}{2} regions. However, the $E(\bv)$ values of \citet{foster15} were obtained from the photometry of point stars considered to be associated with the \ion{H}{2} regions. The Pa$\alpha$-H$\alpha$ $E(\bv)$ values are coincident with those of \citet{foster15} within $\sim$0.4 mag, but a systematic underestimation of the Pa$\alpha$-H$\alpha$ $E(\bv)$ was also found. The offset from the $E(\bv)$ values of \citet{foster15} was estimated to be 0.16 $\pm$ 0.03 mag when a temperature of 10$^{4}$ K for ionized hydrogen gas is assumed. This can be explained by the dust-scattering effect, which is more important at H$\alpha$ than at Pa$\alpha$, when observing emissions from extended sources. We used the observed Pa$\alpha$ total flux and the Pa$\alpha$-H$\alpha$ $E(\bv)$ value to calculate the total Lyman continuum luminosity as a function of distance for each \ion{H}{2} region, which enables the distance to the \ion{H}{2} region and the spectral type(s) of its ionizing star(s) to be constrained. We found that the total luminosities calculated for 9 \ion{H}{2} regions are simultaneously consistent with their known kinematic or spectrophotometric distances and the spectral types of their known ionizing stars.

Using the MIPAPS Pa$\alpha$ and IPHAS H$\alpha$ images, we made an $E(\bv)$ map for the entire region of an \ion{H}{2} region (known as Sh2-131), which is one of the largest \ion{H}{2} regions in the Cepheus region. In the NW part with relatively high extinction, the $E(\bv)$ map shows many high-extinction filamentary features, which suggest the existence of foreground dust clouds in the regions. On the other hand, a central void and the outer rim in the SE part were found to have negative $E(\bv)$ values. This indicates that there is H$\alpha$ excess over that predicted from the intrinsic H$\alpha$ to Pa$\alpha$ ratio of hydrogen recombination lines. The H$\alpha$ excess can be explained by the dust-scattering effect. In fact, in the IPHAS H$\alpha$ image, we can clearly see diffuse and faint H$\alpha$ emission, even in the regions where the Pa$\alpha$ emission is very low or negligible. These diffuse H$\alpha$ emissions are attributable to light scattered by interstellar dust. The radial profile of $E(\bv)$, which decreases rapidly near the SE edge of the source, is consistent with the result expected by the dust-scattering effect.

\acknowledgements{}
We greatly appreciate the efforts of the KASI, SaTReC, and KARI staff in the development and operation of STSAT-3 and MIRIS. We also appreciate the efforts of the ISAS/JAXA and Genesia Co. staff in Japan for their contributions to the MIRIS instrument development. MIRIS was supported by a National Research Foundation of Korea grant funded by the Korean government, and by ISAS/JAXA. This work was also supported by the National Research Foundation of Korea, with grant number NRF-2014M1A3A3A02034746. This paper makes use of data obtained as part of the IPHAS (http://www.iphas.org/) carried out at the Isaac Newton Telescope (INT). We also used the Web site for the {\it WISE} catalog of Galactic \ion{H}{2} regions (http://astro.phys.wvu.edu/wise/). This research made use of Montage. It is funded by the National Science Foundation under Grant Number ACI-1440620, and was previously funded by the National Aeronautics and Space Administration's Earth Science Technology Office, Computation Technologies Project, under Cooperative Agreement Number NCC5-626 between NASA and the California Institute of Technology. This research has made use of the NASA/IPAC Infrared Science Archive, which is operated by the Jet Propulsion Laboratory, California Institute of Technology, under contract with the National Aeronautics and Space Administration. This publication makes use of data products from the Two Micron All Sky Survey, which is a joint project of the University of Massachusetts and the Infrared Processing and Analysis Center/California Institute of Technology, funded by the National Aeronautics and Space Administration and the National Science Foundation.

\clearpage
\begin{landscape}
\LongTables
\begin{deluxetable*}{cccrccrrr}
\tablecaption{{\it WISE} \ion{H}{2} Region Sources: ``Known''\label{table:wk}}
\tablewidth{700pt}
\tabletypesize{\footnotesize}
\tablehead{
\colhead{ID} & \colhead{{\it WISE} Name} & \colhead{MIPAPS Name} & \colhead{Radius} &
\multicolumn{2}{c}{Visual Inspection of Detection} & \colhead{Pa$\alpha$ Total Flux} & \colhead{H$\alpha$ Total Flux} & \colhead{$E(\bv)$} \\
\colhead{} & \colhead{} & \colhead{} & \colhead{(arcsec)} & \colhead{} & \colhead{} &
\colhead{(10$^{-14}$ W m$^{-2}$)} & \colhead{(10$^{-14}$ W m$^{-2}$)} & \colhead{(mag)} \\
\cline{5-6}
\colhead{} & \colhead{} & \colhead{} & \colhead{} & \colhead{Pa$\alpha$\tablenotemark{a} (Stellar Residual Overlap\tablenotemark{b})} & \colhead{H$\alpha$\tablenotemark{a}} & \colhead{} & \colhead{} & \colhead{}
}
\startdata
WK01 & G097.515+03.173 & G097.51+03.17 & 135 & Y (partially) & Y & 2.79 $\pm$ 0.03 & 1.46 $\pm$ 0.01 & 1.46 $\pm$ 0.01 \\
WK02 & G097.528+03.184 & \nodata & \nodata & N (N) & Y & \nodata & \nodata & \nodata \\
WK03 & G099.484+03.801 & G099.36+03.66 & 4535 & Y (partially) & Y & 1166.30 $\pm$ 1.62 & 6705.15 $\pm$ 3.14 & 0.20 $\pm$ 0.00 \\
WK04 & G102.877$-$00.695 & G102.70$-$00.85 & 3085 & Y (partially) & Y & 375.57 $\pm$ 1.01 & 1385.85 $\pm$ 0.62 & 0.44 $\pm$ 0.00 \\
WK05 & G104.546+01.255 & G104.54+01.27 & 1000 & Y (partially) & Y & 31.32 $\pm$ 0.31 & 93.50 $\pm$ 0.12 & 0.55 $\pm$ 0.01 \\
WK06 & G105.779+00.048 & G105.77+00.06 & 740 & Y (partially) & Y & 14.18 $\pm$ 0.22 & 56.81 $\pm$ 0.09 & 0.39 $\pm$ 0.01 \\
WK07 & G106.605+05.252 & G106.61+05.25 & 1145 & Y (partially) & Yp & 40.32 $\pm$ 0.37 & \nodata & \nodata \\
WK08 & G106.809+03.310 & G106.81+03.31 & 260 & Y (N) & Y & 3.02 $\pm$ 0.06 & 2.49 $\pm$ 0.01 & 1.22 $\pm$ 0.01 \\
WK09 & G107.034$-$00.801 & G107.03$-$00.80 & 1600 & Y (partially) & Y & 135.36 $\pm$ 0.47 & 502.16 $\pm$ 0.29  & 0.43 $\pm$ 0.00 \\
WK10 & G107.209$-$01.334 & G107.21$-$01.33 & 300 & Y (partially) & Y & 1.34 $\pm$ 0.07 & 5.97 $\pm$ 0.02 & 0.34 $\pm$ 0.03 \\
WK11 & G108.191+00.586 & G108.19+00.59 & 230 & Y (N) & Y & 10.66 $\pm$ 0.07 & 3.06 $\pm$ 0.02 & 1.77 $\pm$ 0.00 \\
WK12 & G108.273$-$01.066 & G108.28$-$01.07 & 160 & Y (largely) & Y & \nodata & \nodata & \nodata \\
WK13 & G108.375$-$01.056 & G108.37$-$01.06 & 170 & Y (largely) & Y & \nodata & \nodata & \nodata \\
WK14 & G108.503+06.356 & G108.52+06.40 & 2800 & Y (partially) & no data & 127.56 $\pm$ 0.76 & \nodata & \nodata \\
WK15 & G108.752$-$00.972 & \nodata & \nodata & N (N) & N & \nodata & \nodata & \nodata \\
WK16 & G108.758$-$00.989 & \nodata & \nodata & N (N) & N & \nodata & \nodata & \nodata \\
WK17 & G108.764$-$00.952 & G108.76$-$00.95 & 163 & Y (partially) & Y & 10.74 $\pm$ 0.08 & 13.37 $\pm$ 0.02 & 1.00 $\pm$ 0.00 \\
WK18 & G108.770$-$00.974 & \nodata & \nodata & N (N) & N & \nodata & \nodata & \nodata \\
WK19 & G110.099+00.042 & G110.10+00.04 & 250 & Y (N) & Y & 16.41 $\pm$ 0.06 & 16.49 $\pm$ 0.01  & 1.12 $\pm$ 0.00 \\
WK20 & G110.211+02.616 & G110.18+02.52 & 2200 & Y (partially) & Y & 227.14 $\pm$ 0.60 & 290.04 $\pm$ 0.49 & 0.99 $\pm$ 0.00 \\
WK21 & G111.286$-$00.660 & G111.29$-$00.66 & 165 & Y (N) & Y & 6.56 $\pm$ 0.17 & 13.66 $\pm$ 0.05 & 0.74 $\pm$ 0.01 \\
WK22 & G111.558+00.804 & G111.56+00.81 & 375 & Y (partially) & Y & 113.36 $\pm$ 0.19 & 75.80 $\pm$ 0.06 & 1.33 $\pm$ 0.00 \\
WK23 & G111.612+00.371 & G111.61+00.37 & 64 & Y (partially) & Y & \nodata & \nodata & \nodata \\
WK24 & G111.946+01.336 & G111.95+01.29 & 740 & Y (partially) & Y & 25.43 $\pm$ 0.23 & 61.04 $\pm$ 0.12 & 0.66 $\pm$ 0.00 \\
WK25 & G112.212+00.229 & G112.23+00.19 & 1250 & Y (partially) & Y & 107.00 $\pm$ 0.40 & 407.68 $\pm$ 0.29 & 0.42 $\pm$ 0.00 \\
WK26 & G113.595$-$00.749 & G113.59$-$00.77 & 755 & Y (partially) & Y & 24.87 $\pm$ 0.21 & 23.27 $\pm$ 0.07 & 1.15 $\pm$ 0.00 \\
WK27 & G113.900$-$01.613 & G113.92$-$01.61 & 425 & Y (N) & Y & 3.22 $\pm$ 0.11 & 6.02 $\pm$ 0.03 & 0.79 $\pm$ 0.02 \\
WK28 & G114.626+00.219 & G114.60+00.22 & 435 & Y (partially) & Y & 6.69 $\pm$ 0.14 & 17.75 $\pm$ 0.02 & 0.61 $\pm$ 0.01 \\
WK29 & G114.605$-$00.801 & G114.62$-$00.81 & 372 & Y (partially) & Y & 3.17 $\pm$ 0.08 & 5.49 $\pm$ 0.02 & 0.83 $\pm$ 0.01 \\
WK30 & G115.785$-$01.561 & G115.79$-$01.57 & 326 & Y (partially) & Y & 12.85 $\pm$ 0.11 & 23.48 $\pm$ 0.02 & 0.80 $\pm$ 0.00 \\
WK31 & G115.885$-$01.707 & G115.89$-$01.71 & 285 & Y (N) & Y & 1.04 $\pm$ 0.11 & 2.57 $\pm$ 0.02 & 0.65 $\pm$ 0.05 \\
\enddata
\tablenotetext{a}{``Y'': visually detected; ``Yp'': visually detected but partially observed; ``N'': not visually detected; ``no data'': not observed.}
\tablenotetext{b}{``N'': not overlapped with any stellar residuals; ``partially'': partially overlapped with stellar residuals; ``largely'': largely overlapped with stellar residuals.}
\tablecomments{MIPAPS names are assigned to the sources with Pa$\alpha$ detections according to the central positions of the individual Pa$\alpha$ features. Radius values approximate the angular extents of the Pa$\alpha$ features, and they were used as aperture radii for the photometries of the Pa$\alpha$ and H$\alpha$ total fluxes. The red solid circles with these central positions and radii are denoted in Figures \ref{fig:paa} and \ref{fig:ha}.
}
\end{deluxetable*}

\LongTables
\begin{deluxetable*}{cccrccrrr}
\tablecaption{{\it WISE} \ion{H}{2} Region Sources: ``Candidate''\label{table:wc}}
\tablewidth{700pt}
\tabletypesize{\footnotesize}
\tablehead{
\colhead{ID} & \colhead{{\it WISE} Name} & \colhead{MIPAPS Name} & \colhead{Radius} &
\multicolumn{2}{c}{Visual Inspection of Detection} & \colhead{Pa$\alpha$ Total Flux} & \colhead{H$\alpha$ Total Flux} & \colhead{$E(\bv)$} \\
\colhead{} & \colhead{} & \colhead{} & \colhead{(arcsec)} & \colhead{} & \colhead{} &
\colhead{(10$^{-14}$ W m$^{-2}$)} & \colhead{(10$^{-14}$ W m$^{-2}$)} & \colhead{(mag)} \\
\cline{5-6}
\colhead{} & \colhead{} & \colhead{} & \colhead{} & \colhead{Pa$\alpha$\tablenotemark{a} (Stellar Residual Overlap\tablenotemark{b})} & \colhead{H$\alpha$\tablenotemark{a}} & \colhead{} & \colhead{} & \colhead{}
}
\startdata
WC01 & G097.210+03.245 & \nodata & \nodata & N (N) & N & \nodata & \nodata & \nodata \\
WC02 & G097.252+03.320 & G097.25+03.33 & 190 & Y (N) & Y & 0.72 $\pm$ 0.06 & 0.13 $\pm$ 0.01 & 2.01 $\pm$ 0.06 \\
WC03 & G097.311+03.269\tablenotemark{c} & G097.31+03.27 & 110 & Y (partially) & Y & 0.22 $\pm$ 0.02 & 0.04 $\pm$ 0.01 & 2.04 $\pm$ 0.10 \\
WC04 & G097.444+03.083\tablenotemark{c} & G097.44+03.08 & 120 & Y (partially) & Y & 0.23 $\pm$ 0.04 & 0.37 $\pm$ 0.00 & 0.87 $\pm$ 0.08 \\
WC05 & G097.728+02.352 & \nodata & \nodata & N (N) & Y & \nodata & \nodata & \nodata \\
WC06 & G098.320+01.552 & \nodata & \nodata & N (largely) & N & \nodata & \nodata & \nodata \\
WC07 & G098.855+02.933 & \nodata & \nodata & N (N) & Y & \nodata & \nodata & \nodata \\
WC08 & G100.169+02.026 & \nodata & \nodata & N (entirely) & N & \nodata & \nodata & \nodata \\
WC09 & G100.181+02.038 & \nodata & \nodata & N (entirely) & Y & \nodata & \nodata & \nodata \\
WC10 & G100.205+01.885\tablenotemark{c} & G100.20+01.89 & 130 & Y (partially) & Y & 0.16 $\pm$ 0.02 & 0.02 $\pm$ 0.01 & 2.08 $\pm$ 0.20 \\
WC11 & G100.199+02.064 & G100.20+02.07 & 100 & Y (largely) & Y & \nodata & \nodata & \nodata \\
WC12 & G100.714$-$00.527 & G100.71$-$00.50 & 500 & Y (partially) & Y & 2.46 $\pm$ 0.13 & 5.86 $\pm$ 0.06 & 0.67 $\pm$ 0.03 \\
WC13 & G101.016+02.590\tablenotemark{c} & \nodata & \nodata & N (N) & Y & \nodata & \nodata & \nodata \\
WC14 & G101.065+02.499\tablenotemark{c} & G101.08+02.48 & 1060 & Y (partially) & Y & 32.07 $\pm$ 0.28 & 33.67 $\pm$ 0.11 & 1.09 $\pm$ 0.00 \\
WC15 & G101.439+02.653\tablenotemark{c} & G101.44+02.65 & 220 & Y (N) & Y & 1.25 $\pm$ 0.06 & 0.39 $\pm$ 0.01 & 1.73 $\pm$ 0.03 \\
WC16 & G101.527$-$00.515 & \nodata & \nodata & N (partially) & Y & \nodata & \nodata & \nodata \\
WC17 & G101.663+02.820 & \nodata & \nodata & N (N) & N & \nodata & \nodata & \nodata \\
WC18 & G101.763+02.808 & \nodata & \nodata & N (N) & N & \nodata & \nodata & \nodata \\
WC19 & G102.051+02.861 & \nodata & \nodata & N (partially) & N & \nodata & \nodata & \nodata \\
WC20 & G102.207$-$00.736 & \nodata & \nodata & N (N) & Y & \nodata & \nodata & \nodata \\
WC21 & G102.327+03.681 & \nodata & \nodata & N (N) & Y & \nodata & \nodata & \nodata \\
WC22 & G102.354+03.635 & \nodata & \nodata & N (N) & Y & \nodata & \nodata & \nodata \\
WC23 & G102.807+01.204 & G102.81+01.20 & 190 & Y (partially) & Y & 0.14 $\pm$ 0.04 & 0.10 $\pm$ 0.01 & 1.28 $\pm$ 0.16 \\
WC24 & G103.578+03.165 & G103.62+03.18 & 410 & Y (partially) & Y & 0.84 $\pm$ 0.13 & 1.70 $\pm$ 0.04 & 0.75 $\pm$ 0.08 \\
WC25 & G103.659+02.151 & G103.65+02.15 & 110 & Y (N) & Y & 0.28 $\pm$ 0.04 & 0.16 $\pm$ 0.01 & 1.42 $\pm$ 0.08 \\
WC26 & G103.686+00.425\tablenotemark{c} & G103.69+00.41 & 290 & Y (largely) & Y & \nodata & \nodata & \nodata \\
WC27 & G103.743+02.162\tablenotemark{c} & G103.74+02.15 & 300 & Y (N) & Y & 4.30 $\pm$ 0.11 & 5.12 $\pm$ 0.03 & 1.03 $\pm$ 0.01 \\
WC28 & G103.875+01.857 & \nodata & \nodata & N (partially) & N & \nodata & \nodata & \nodata \\
WC29 & G104.153+00.259 & G104.14+00.26 & 1030 & Y (partially) & Y & 2.32 $\pm$ 0.31 & 15.16 $\pm$ 0.10 & 0.14 $\pm$ 0.07 \\
WC30 & G104.355+00.404 & \nodata & \nodata & N (N) & Y & \nodata & \nodata & \nodata \\
WC31 & G104.393+00.049 & G104.39+00.05 & 270 & Y (partially) & Y & 0.77 $\pm$ 0.07 & 1.50 $\pm$ 0.02 & 0.77 $\pm$ 0.05 \\
WC32 & G104.648+00.110 & G104.65+00.10 & 850 & Y (partially) & Y & 3.32 $\pm$ 0.24 & 19.65 $\pm$ 0.10 & 0.19 $\pm$ 0.04 \\
WC33 & G104.675+00.595 & G104.67+00.61 & 280 & Y (partially) & Y & 1.26 $\pm$ 0.07 & 0.72 $\pm$ 0.02 & 1.41 $\pm$ 0.03 \\
WC34 & G104.700+02.784\tablenotemark{c} & \nodata & \nodata & N (N) & Y & \nodata & \nodata & \nodata \\
WC35 & G104.716+02.813\tablenotemark{c} & G104.71+02.83 & 350 & Y (partially) & Y & 4.28 $\pm$ 0.10 & 1.35 $\pm$ 0.03 & 1.72 $\pm$ 0.02 \\
WC36 & G104.728+00.446 & \nodata & \nodata & N (largely) & N & \nodata & \nodata & \nodata \\
WC37 & G105.307+04.058 & \nodata & \nodata & N (partially) & N & \nodata & \nodata & \nodata \\
WC38 & G105.365+03.234 & G105.37+03.22 & 700 & Y (partially) & Y & 2.75 $\pm$ 0.19 & 12.85 $\pm$ 0.06 & 0.31 $\pm$ 0.04 \\
WC39 & G105.487+03.889 & \nodata & \nodata & N (N) & Y & \nodata & \nodata & \nodata \\
WC40 & G105.571+00.296 & G105.57+00.30 & 140 & Y (partially) & Y & 0.58 $\pm$ 0.05 & 0.46 $\pm$ 0.02 & 1.24 $\pm$ 0.05 \\
WC41 & G105.635+00.345\tablenotemark{c} & G105.63+00.34 & 150& Y (largely) & Y & \nodata & \nodata & \nodata \\
WC42 & G105.744+00.298 & G105.75+00.30 & 580 & Y (partially) & Y & 7.13 $\pm$ 0.19 & 22.88 $\pm$ 0.09 & 0.51 $\pm$ 0.01 \\
WC43 & G105.852+00.142 & G105.85+00.14 & 100 & Y (N) & Y & \nodata & \nodata & \nodata \\
WC44 & G106.241+00.957\tablenotemark{c} & G106.24+00.97 & 170 & Y (partially) & Y & 0.33 $\pm$ 0.04 & 0.05 $\pm$ 0.01 & 2.15 $\pm$ 0.11 \\
WC45 & G106.499+00.925 & \nodata & \nodata & N (partially) & N & \nodata & \nodata & \nodata \\
WC46 & G108.213$-$01.293 & \nodata & \nodata & N (N) & Y & \nodata & \nodata & \nodata \\
WC47 & G108.902+02.714 & G108.88+02.73 & 340 & Y (partially) & Yp & 0.86 $\pm$ 0.10 & \nodata & \nodata \\
WC48 & G109.068$-$00.322\tablenotemark{c} & G109.08$-$00.32 & 160 & Y (N) & Y & \nodata & \nodata & \nodata \\
WC49 & G109.104$-$00.347\tablenotemark{c} & G109.10$-$00.34 & 170 & Y (N) & Y & 1.77 $\pm$ 0.08 & 1.64 $\pm$ 0.02 & 1.16 $\pm$  0.02 \\
WC50 & G109.854+02.147 & \nodata & \nodata & N (entirely) & N & \nodata & \nodata & \nodata \\
WC51 & G109.874+02.115 & G109.87+02.11 & 160 & Y (partially) & Y & 1.04 $\pm$ 0.07 & 0.25 $\pm$ 0.00 & 1.86 $\pm$ 0.04 \\
WC52 & G109.927+01.981\tablenotemark{c} & G109.92+01.98 & 170 & Y (N) & Y & 0.27 $\pm$ 0.05 & 0.37 $\pm$ 0.01 & 0.95 $\pm$ 0.10 \\
WC53 & G109.994$-$00.092 & \nodata & \nodata & N (N) & Y & \nodata & \nodata & \nodata \\
WC54 & G110.252+00.009\tablenotemark{c} & G110.25+00.01 & 100 & Y (partially) & Y & 0.39 $\pm$ 0.03 & 0.44 $\pm$ 0.00 & 1.05 $\pm$ 0.04 \\
WC55 & G110.286+02.488\tablenotemark{c} & G110.29+02.49 & 95 & Y (N) & N & 0.09 $\pm$ 0.04 & \nodata & \nodata \\
WC56 & G110.923+01.906 & G110.92+01.89 & 530 & Y (partially) & Y & 2.93 $\pm$ 0.14 & 5.15 $\pm$ 0.03 & 0.82 $\pm$ 0.03 \\
WC57 & G111.329+00.783 & \nodata & \nodata & N (N) & N & \nodata & \nodata & \nodata \\
WC58 & G111.653+00.950\tablenotemark{c} & G111.63+00.98 & 350 & Y (partially) & Y & 7.54 $\pm$ 0.15 & 9.47 $\pm$ 0.05 & 1.00 $\pm$ 0.01 \\
WC59 & G111.802+00.526\tablenotemark{c} & G111.80+00.53 & 110 & Y (N) & Y & 0.22 $\pm$ 0.03 & 0.08 $\pm$ 0.01 & 1.65 $\pm$ 0.09 \\
WC60 & G111.873+00.820 & \nodata & \nodata & N (entirely) & N & \nodata & \nodata & \nodata \\
WC61 & G111.907+00.800 & \nodata & \nodata & N (partially) & Y & \nodata & \nodata & \nodata \\
WC62 & G111.922+00.859 & \nodata & \nodata & N (N) & N & \nodata & \nodata & \nodata \\
WC63 & G111.947+00.799 & \nodata & \nodata & N (partially) & N & \nodata & \nodata & \nodata \\
WC64 & G111.966+00.839 & G111.96+00.84 & 100 & Y (partially) & Y & 0.14 $\pm$ 0.02 & 0.12 $\pm$ 0.01 & 1.21 $\pm$ 0.08 \\
WC65 & G113.009$-$01.393 & G113.00$-$01.37 & 2015 & Y (partially) & Y & 28.54 $\pm$ 0.63 & \nodata & \nodata \\
WC66 & G113.096+02.602 & \nodata & \nodata & N (N) & N & \nodata & \nodata & \nodata \\
WC67 & G114.312$-$00.510 & G114.29$-$00.51 & 520 & Y (partially) & Y & 1.70 $\pm$ 0.13 & 2.98 $\pm$ 0.03 & 0.82 $\pm$ 0.04 \\
WC68 & G114.332+00.788 & \nodata & \nodata & N (N) & N & \nodata & \nodata & \nodata \\
WC69 & G114.473$-$00.430 & \nodata & \nodata & N (N) & Y & \nodata & \nodata & \nodata \\
WC70 & G114.526$-$00.543\tablenotemark{c} & G114.53$-$00.54 & 220 & Y (partially) & Y & 0.63 $\pm$ 0.05 & 0.16 $\pm$ 0.01 & 1.85 $\pm$ 0.05 \\
WC71 & G115.105$-$01.438 & G115.11$-$01.44 & 120 & Y (N) & Y & 0.11 $\pm$ 0.02 & \nodata & \nodata \\
\enddata
\tablenotetext{a}{``Y'': visually detected; ``Yp'': visually detected but partially observed; ``N'': not visually detected.}
\tablenotetext{b}{``N'': not overlapped with any stellar residuals; ``partially'': partially overlapped with stellar residuals; ``largely'': largely overlapped with stellar residuals; ``entirely'': entirely overlapped with stellar residuals.}
\tablenotetext{c}{These sources were updated to be ``Known'' sources in the most up-to-date version 2.0 of the {\it WISE} \ion{H}{2} Region catalog (http://astro.phys.wvu.edu/wise/).}
\tablecomments{MIPAPS names are assigned to the sources with Pa$\alpha$ detections according to the central positions of the individual Pa$\alpha$ features. Radius values approximate the angular extents of the Pa$\alpha$ features, and they were used as aperture radii for the photometries of the Pa$\alpha$ and H$\alpha$ total fluxes. The cyan solid circles with these central positions and radii are denoted in Figures \ref{fig:paa} and \ref{fig:ha}.
}
\end{deluxetable*}

\LongTables
\begin{deluxetable*}{cccrccrrr}
\tablecaption{{\it WISE} \ion{H}{2} Region Sources: ``Group''\label{table:wg}}
\tablewidth{700pt}
\tabletypesize{\footnotesize}
\tablehead{
\colhead{ID} & \colhead{{\it WISE} Name} & \colhead{MIPAPS Name} & \colhead{Radius} &
\multicolumn{2}{c}{Visual Inspection of Detection} & \colhead{Pa$\alpha$ Total Flux} & \colhead{H$\alpha$ Total Flux} & \colhead{$E(\bv)$} \\
\colhead{} & \colhead{} & \colhead{} & \colhead{(arcsec)} & \colhead{} & \colhead{} &
\colhead{(10$^{-14}$ W m$^{-2}$)} & \colhead{(10$^{-14}$ W m$^{-2}$)} & \colhead{(mag)} \\
\cline{5-6}
\colhead{} & \colhead{} & \colhead{} & \colhead{} & \colhead{Pa$\alpha$\tablenotemark{a} (Stellar Residual Overlap\tablenotemark{b})} & \colhead{H$\alpha$\tablenotemark{a}} & \colhead{} & \colhead{} & \colhead{}
}
\startdata
WG01 & G099.091+02.969 & G099.09+02.97 & 113 & Y (N) & Y & 0.22 $\pm$ 0.04 & 0.92 $\pm$ 0.03 & 0.37 $\pm$ 0.10 \\
WG02 & G103.061$-$00.691 & G103.06$-$00.69 & 110 & Y (N) & Y & 2.21 $\pm$ 0.31 & 1.13 $\pm$ 0.08 & 1.47 $\pm$ 0.08 \\
WG03 & G106.798+05.313 & G106.80+05.31 & 115 & Y (largely) & no data & \nodata & \nodata & \nodata \\
WG04 & G107.333+05.061\tablenotemark{c} & G107.63+04.96 & 2500 & Y (partially) & Yp & 236.30 $\pm$ 0.88 & \nodata & \nodata \\
WG05 & G107.866+05.607 & G107.87+05.61 & 1968 & Y (partially) & no data & 94.60 $\pm$ 1.08 & \nodata & \nodata \\
WG06 & G110.927$-$00.731 & G110.95$-$00.73 & 1080 & Y (partially) & Y & 121.69 $\pm$ 0.71 & 312.75 $\pm$ 0.23 & 0.63 $\pm$ 0.00 \\
WG07 & G111.196$-$00.798 & \nodata & \nodata & N (partially) & Y & \nodata & \nodata & \nodata \\
WG08 & G111.245$-$00.924 & G111.24$-$00.94 & 275 & Y (N) & Y & 5.36 $\pm$ 0.26 & 7.71 $\pm$ 0.05 & 0.93 $\pm$ 0.03 \\
WG09 & G111.430$-$00.790 & G111.35$-$00.84 & 365 & Y (partially) & Y & 10.50 $\pm$ 0.45 & 23.83 $\pm$ 0.11 & 0.69 $\pm$ 0.02 \\
WG10 & G111.478$-$00.591 & G111.44$-$00.65 & 455 & Y (partially) & Y & 38.09 $\pm$ 0.42 & 63.18 $\pm$ 0.12 & 0.85 $\pm$ 0.01 \\
WG11 & G111.543+00.775 & \nodata & \nodata & N (N) & Y & \nodata & \nodata & \nodata \\
WG12 & G111.601+00.393 & \nodata & \nodata & N (partially) & Y & \nodata & \nodata & \nodata \\
WG13 & G111.640+00.360 & G111.64+00.37 & 320 & Y (partially) & Y & 12.28 $\pm$ 0.09 & 8.86 $\pm$ 0.03 & 1.29 $\pm$ 0.00 \\
WG14 & G112.071+01.063\tablenotemark{c} & G112.06+01.02 & 527 & Y (partially) & Y & 10.93 $\pm$ 0.19 & 18.16 $\pm$ 0.06 & 0.85 $\pm$ 0.01 \\
WG15 & G113.614$-$00.615 & \nodata & \nodata & N (N) & Y & \nodata & \nodata & \nodata \\
WG16 & G114.437$-$00.662 & G114.43$-$00.65 & 214 & Y (partially) & Y & 0.29 $\pm$ 0.04 & 0.28 $\pm$ 0.01 & 1.13 $\pm$ 0.08 \\
WG17 & G114.496$-$00.946 & \nodata & \nodata & N (largely) & Y & \nodata & \nodata & \nodata \\
WG18 & G114.520$-$00.867 & \nodata & \nodata & N (N) & Y & \nodata & \nodata & \nodata \\
\enddata
\tablenotetext{a}{``Y'': visually detected; ``Yp'': visually detected but partially observed; ``N'': not visually detected; ``no data'': not observed.}
\tablenotetext{b}{``N'': not overlapped with any stellar residuals; ``partially'': partially overlapped with stellar residuals; ``largely'': largely overlapped with stellar residuals.}
\tablenotetext{c}{These sources were updated to be ``Known'' sources in the most up-to-date version 2.0 of the {\it WISE} \ion{H}{2} Region catalog (http://astro.phys.wvu.edu/wise/).}
\tablecomments{MIPAPS names are assigned to the sources with Pa$\alpha$ detections according to the central positions of the individual Pa$\alpha$ features. Radius values approximate the angular extents of the Pa$\alpha$ features, and they were used as aperture radii for the photometries of the Pa$\alpha$ and H$\alpha$ total fluxes. The green solid circles with these central positions and radii are denoted in Figures \ref{fig:paa} and \ref{fig:ha}.
}
\end{deluxetable*}

\LongTables
\begin{deluxetable*}{cccrccrrr}
\tablecaption{{\it WISE} \ion{H}{2} Region Sources: ``Radio Quiet''\label{table:wr}}
\tablewidth{700pt}
\tabletypesize{\footnotesize}
\tablehead{
\colhead{ID} & \colhead{{\it WISE} Name} & \colhead{MIPAPS Name} & \colhead{Radius} &
\multicolumn{2}{c}{Visual Inspection of Detection} & \colhead{Pa$\alpha$ Total Flux} & \colhead{H$\alpha$ Total Flux} & \colhead{$E(\bv)$} \\
\colhead{} & \colhead{} & \colhead{} & \colhead{(arcsec)} & \colhead{} & \colhead{} &
\colhead{(10$^{-14}$ W m$^{-2}$)} & \colhead{(10$^{-14}$ W m$^{-2}$)} & \colhead{(mag)} \\
\cline{5-6}
\colhead{} & \colhead{} & \colhead{} & \colhead{} & \colhead{Pa$\alpha$\tablenotemark{a} (Stellar Residual Overlap\tablenotemark{b})} & \colhead{H$\alpha$\tablenotemark{a}} & \colhead{} & \colhead{} & \colhead{}
}
\startdata
WR01 & G097.186+03.768 & \nodata & \nodata & N (N) & N & \nodata & \nodata & \nodata \\
WR02 & G097.240+03.300 & \nodata & \nodata & N (N) & N & \nodata & \nodata & \nodata \\
WR03 & G097.276+03.298 & \nodata & \nodata & N (N) & N & \nodata & \nodata & \nodata \\
WR04 & G097.284+03.179 & \nodata & \nodata & N (N) & N & \nodata & \nodata & \nodata \\
WR05 & G097.325+03.226 & \nodata & \nodata & N (largely) & N & \nodata & \nodata & \nodata \\
WR06 & G097.701+03.821 & \nodata & \nodata & N (N) & N & \nodata & \nodata & \nodata \\
WR07 & G098.908+02.716 & \nodata & \nodata & N (N) & N & \nodata & \nodata & \nodata \\
WR08 & G099.041+02.568 & \nodata & \nodata & N (largely) & N & \nodata & \nodata & \nodata \\
WR09 & G100.015+02.359 & \nodata & \nodata & N (N) & Y & \nodata & \nodata & \nodata \\
WR10 & G100.126+02.373 & \nodata & \nodata & N (N) & N & \nodata & \nodata & \nodata \\
WR11 & G100.161+01.766 & \nodata & \nodata & N (largely) & N & \nodata & \nodata & \nodata \\
WR12 & G100.162+01.665 & \nodata & \nodata & N (largely) & N & \nodata & \nodata & \nodata \\
WR13 & G100.168+02.082 & \nodata & \nodata & N (largely) & N & \nodata & \nodata & \nodata \\
WR14 & G100.205+01.824 & \nodata & \nodata & N (N) & N & \nodata & \nodata & \nodata \\
WR15 & G100.213+01.883 & \nodata & \nodata & N (N) & Y & \nodata & \nodata & \nodata \\
WR16 & G100.225+02.031 & \nodata & \nodata & N (largely) & N & \nodata & \nodata & \nodata \\
WR17 & G100.238+01.633 & \nodata & \nodata & N (largely) & N & \nodata & \nodata & \nodata \\
WR18 & G100.338+01.691 & \nodata & \nodata & N (largely) & N & \nodata & \nodata & \nodata \\
WR19 & G100.479+00.190 & \nodata & \nodata & N (N) & N & \nodata & \nodata & \nodata \\
WR20 & G100.479$-$00.637 & \nodata & \nodata & N (N) & N & \nodata & \nodata & \nodata \\
WR21 & G100.879+00.662 & \nodata & \nodata & N (partially) & N & \nodata & \nodata & \nodata \\
WR22 & G101.520$-$00.453 & \nodata & \nodata & N (partially) & Y & \nodata & \nodata & \nodata \\
WR23 & G102.120+03.565 & \nodata & \nodata & N (partially) & N & \nodata & \nodata & \nodata \\
WR24 & G102.311+03.677 & \nodata & \nodata & N (N) & Y & \nodata & \nodata & \nodata \\
WR25 & G102.334+03.608 & \nodata & \nodata & N (N) & Y & \nodata & \nodata & \nodata \\
WR26 & G102.589$-$00.776 & \nodata & \nodata & N (largely) & N & \nodata & \nodata & \nodata \\
WR27 & G102.631+03.757 & \nodata & \nodata & N (partially) & N & \nodata & \nodata & \nodata \\
WR28 & G103.459+01.449 & \nodata & \nodata & N (N) & N & \nodata & \nodata & \nodata \\
WR29 & G103.483+01.998 & \nodata & \nodata & N (N) & N & \nodata & \nodata & \nodata \\
WR30 & G103.640+01.087 & \nodata & \nodata & N (entirely) & N & \nodata & \nodata & \nodata \\
WR31 & G103.690+00.434 & \nodata & \nodata & N (entirely) & N & \nodata & \nodata & \nodata \\
WR32 & G103.745+02.182 & \nodata & \nodata & N (N) & N & \nodata & \nodata & \nodata \\
WR33 & G103.804+00.405 & \nodata & \nodata & N (N) & N & \nodata & \nodata & \nodata \\
WR34 & G103.954+01.097 & \nodata & \nodata & N (partially) & N & \nodata & \nodata & \nodata \\
WR35 & G103.972$-$00.129 & \nodata & \nodata & N (partially) & N & \nodata & \nodata & \nodata \\
WR36 & G105.141+03.415 & \nodata & \nodata & N (largely) & N & \nodata & \nodata & \nodata \\
WR37 & G105.227$-$00.262 & \nodata & \nodata & N (partially) & N & \nodata & \nodata & \nodata \\
WR38 & G105.279$-$00.113 & \nodata & \nodata & N (partially) & Y & \nodata & \nodata & \nodata \\
WR39 & G105.509+00.230 & \nodata & \nodata & N (N) & N & \nodata & \nodata & \nodata \\
WR40 & G105.880+04.253 & \nodata & \nodata & N (N) & N & \nodata & \nodata & \nodata \\
WR41 & G105.962+00.420 & \nodata & \nodata & N (largely) & N & \nodata & \nodata & \nodata \\
WR42 & G106.142+00.129 & \nodata & \nodata & N (N) & N & \nodata & \nodata & \nodata \\
WR43 & G106.909+03.148 & \nodata & \nodata & N (N) & Y & \nodata & \nodata & \nodata \\
WR44 & G107.156$-$00.988 & \nodata & \nodata & N (N) & N & \nodata & \nodata & \nodata \\
WR45 & G107.298+05.638 & \nodata & \nodata & N (partially) & no data & \nodata & \nodata & \nodata \\
WR46 & G107.678+00.235 & \nodata & \nodata & N (N) & N & \nodata & \nodata & \nodata \\
WR47 & G107.683$-$02.239 & \nodata & \nodata & N (N) & Y & \nodata & \nodata & \nodata \\
WR48 & G108.184+05.518 & \nodata & \nodata & N (N) & no data & \nodata & \nodata & \nodata \\
WR49 & G108.394$-$01.046 & \nodata & \nodata & N (entirely) & Y & \nodata & \nodata & \nodata \\
WR50 & G108.412$-$01.097 & \nodata & \nodata & N (partially) & N & \nodata & \nodata & \nodata \\
WR51 & G108.603+00.494 & \nodata & \nodata & N (partially) & N & \nodata & \nodata & \nodata \\
WR52 & G108.666$-$00.391 & \nodata & \nodata & N (partially) & N & \nodata & \nodata & \nodata \\
WR53 & G108.966+02.726 & \nodata & \nodata & N (N) & no data & \nodata & \nodata & \nodata \\
WR54 & G109.285$-$00.987 & \nodata & \nodata & N (N) & N & \nodata & \nodata & \nodata \\
WR55 & G109.621+02.312 & \nodata & \nodata & N (partially) & Y & \nodata & \nodata & \nodata \\
WR56 & G109.878+04.261 & \nodata & \nodata & N (N) & N & \nodata & \nodata & \nodata \\
WR57 & G109.919+00.813 & \nodata & \nodata & N (N) & no data & \nodata & \nodata & \nodata \\
WR58 & G110.016+00.259 & \nodata & \nodata & N (partially) & N & \nodata & \nodata & \nodata \\
WR59 & G110.054$-$00.107 & \nodata & \nodata & N (N) & N & \nodata & \nodata & \nodata \\
WR60 & G110.094$-$00.064 & \nodata & \nodata & N (N) & Y & \nodata & \nodata & \nodata \\
WR61 & G110.135$-$00.077 & \nodata & \nodata & N (N) & N & \nodata & \nodata & \nodata \\
WR62 & G110.160+00.040 & \nodata & \nodata & N (largely) & N & \nodata & \nodata & \nodata \\
WR63 & G110.170+02.630 & \nodata & \nodata & N (entirely) & N & \nodata & \nodata & \nodata \\
WR64 & G110.199+00.016 & \nodata & \nodata & N (entirely) & Y & \nodata & \nodata & \nodata \\
WR65 & G110.505$-$00.586 & \nodata & \nodata & N (N) & N & \nodata & \nodata & \nodata \\
WR66 & G110.548+02.622 & \nodata & \nodata & N (entirely) & Y & \nodata & \nodata & \nodata \\
WR67 & G110.812$-$00.799 & \nodata & \nodata & N (N) & N & \nodata & \nodata & \nodata \\
WR68 & G110.941+01.018 & \nodata & \nodata & N (partially) & N & \nodata & \nodata & \nodata \\
WR69 & G111.046+01.085 & \nodata & \nodata & N (N) & N & \nodata & \nodata & \nodata \\
WR70 & G111.125$-$00.757 & \nodata & \nodata & N (N) & N & \nodata & \nodata & \nodata \\
WR71 & G111.180$-$02.419 & \nodata & \nodata & N (partially) & Y & \nodata & \nodata & \nodata \\
WR72 & G111.236$-$01.238 & G111.23$-$01.23 & 128 & Y (N) & Y & 0.44 $\pm$ 0.06 & 0.54 $\pm$ 0.03 & 1.01 $\pm$ 0.08 \\
WR73 & G111.498+00.369 & G111.49+00.38 & 300 & Y (largely) & Y & \nodata & \nodata & \nodata \\
WR74 & G111.567+00.751 & \nodata & \nodata & N (N) & N & \nodata & \nodata & \nodata \\
WR75 & G111.670+03.264 & \nodata & \nodata & N (N) & N & \nodata & \nodata & \nodata \\
WR76 & G111.774+00.689 & \nodata & \nodata & N (N) & N & \nodata & \nodata & \nodata \\
WR77 & G111.860+00.800 & \nodata & \nodata & N (largely) & N & \nodata & \nodata & \nodata \\
WR78 & G111.861+01.001 & \nodata & \nodata & N (N) & N & \nodata & \nodata & \nodata \\
WR79 & G111.870+00.881 & \nodata & \nodata & N (N) & N & \nodata & \nodata & \nodata \\
WR80 & G111.893+00.991 & \nodata & \nodata & N (N) & N & \nodata & \nodata & \nodata \\
WR81 & G111.941+00.677 & \nodata & \nodata & N (N) & N & \nodata & \nodata & \nodata \\
WR82 & G112.434+00.034 & \nodata & \nodata & N (N) & N & \nodata & \nodata & \nodata \\
WR83 & G112.970$-$00.608 & \nodata & \nodata & N (N) & Y & \nodata & \nodata & \nodata \\
WR84 & G113.246+00.511 & \nodata & \nodata & N (N) & N & \nodata & \nodata & \nodata \\
WR85 & G113.284+00.558 & \nodata & \nodata & N (N) & N & \nodata & \nodata & \nodata \\
WR86 & G113.289$-$00.819 & \nodata & \nodata & N (partially) & N & \nodata & \nodata & \nodata \\
WR87 & G113.566$-$00.698 & \nodata & \nodata & N (N) & N & \nodata & \nodata & \nodata \\
WR88 & G113.569$-$00.657 & \nodata & \nodata & N (N) & N & \nodata & \nodata & \nodata \\
WR89 & G114.082$-$01.401 & \nodata & \nodata & N (entirely) & Y & \nodata & \nodata & \nodata \\
WR90 & G114.569+00.290 & \nodata & \nodata & N (largely) & N & \nodata & \nodata & \nodata \\
WR91 & G114.677$-$02.208 & \nodata & \nodata & N (N) & N & \nodata & \nodata & \nodata \\
WR92 & G115.782$-$01.689 & \nodata & \nodata & N (N) & N & \nodata & \nodata & \nodata \\
\enddata
\tablenotetext{a}{``Y'': visually detected; ``N'': not visually detected; ``no data'': not observed.}
\tablenotetext{b}{``N'': not overlapped with any stellar residuals; ``partially'': partially overlapped with stellar residuals; ``largely'': largely overlapped with stellar residuals; ``entirely'': entirely overlapped with stellar residuals.}
\tablecomments{MIPAPS names are assigned to the sources with Pa$\alpha$ detections according to the central positions of the individual Pa$\alpha$ features. Radius values approximate the angular extents of the Pa$\alpha$ features, and they were used as aperture radii for the photometries of the Pa$\alpha$ and H$\alpha$ total fluxes. The orange solid circles with these central positions and radii are denoted in Figures \ref{fig:paa} and \ref{fig:ha}.
}
\end{deluxetable*}

\clearpage
\begin{deluxetable*}{ccrrlcrrr}
\tablecaption{MIPAPS Pa$\alpha$ Extended Sources\label{table:mpe}}
\tablewidth{700pt}
\tabletypesize{\footnotesize}
\tablehead{
\colhead{ID} & \colhead{MIPAPS Name} & \colhead{Radius} & \colhead{Position Angle} & \colhead{Corresponding Known \ion{H}{2} Region} &
\colhead{IPHAS H$\alpha$ Detection\tablenotemark{a}} & \colhead{Pa$\alpha$ Total Flux} & \colhead{H$\alpha$ Total Flux} & \colhead{$E(\bv)$} \\
\colhead{} & \colhead{} & \colhead{(arcsec)} & \colhead{(deg)} & \colhead{} &
\colhead{} & \colhead{(10$^{-14}$ W m$^{-2}$)} & \colhead{(10$^{-14}$ W m$^{-2}$)} & \colhead{(mag)}
}
\startdata
MPE01 & G099.14+05.26 & 1900 & \nodata & \nodata & Yp & 46.87 $\pm$ 0.67 & \nodata & \nodata \\
MPE02 & G100.64+02.58 & 485 & \nodata & \nodata & Y & 1.53 $\pm$ 0.12 & 4.15 $\pm$ 0.04 & 0.60 $\pm$ 0.04 \\
MPE03 & G103.45$-$02.29 & 1380 & \nodata & GAL 103.39$-$02.28 & Y & 18.76 $\pm$ 0.41 & 52.15 $\pm$ 0.17 & 0.58 $\pm$ 0.01 \\
MPE04 & G103.82+02.61 & 4800 & \nodata & Sh2-134 & Y & \nodata & \nodata & \nodata \\
MPE05 & G104.71+04.45 & 1300 & \nodata & LBN 494 & Y & 13.54 $\pm$ 0.41 & 93.41 $\pm$ 0.19 & 0.11 $\pm$ 0.02 \\
MPE06 & G105.39+01.97 & 1900 & \nodata & [C51] 93 & Y & 42.10 $\pm$ 0.66 & 316.71 $\pm$ 0.26 & 0.06 $\pm$ 0.01 \\
MPE07 & G106.20$-$01.00 & 1460; 890 & 85 & LBN 506 & Y & \nodata & \nodata & \nodata \\
MPE08 & G106.35$-$01.78 & 880 & \nodata & \nodata & Y & 3.46 $\pm$ 0.22 & 15.71 $\pm$ 0.10 & 0.33 $\pm$ 0.03 \\
MPE09 & G107.63+00.06 & 930 & \nodata & \nodata & Y & 7.50 $\pm$ 0.25 & 9.45 $\pm$ 0.06 & 1.00 $\pm$ 0.02 \\
MPE10 & G107.89$-$01.87 & 530 & \nodata & \nodata & Y & 0.82 $\pm$ 0.12 & 5.41 $\pm$ 0.03 & 0.13 $\pm$ 0.08 \\
MPE11 & G107.99$-$00.33 & 1070 & \nodata & LBN 513; Du 53; BFS 12; BFS 13 & Y & 14.42 $\pm$ 0.30 & 37.85 $\pm$ 0.08 & 0.61 $\pm$ 0.01 \\
MPE12 & G108.44$-$01.95 & 800 & \nodata & \nodata & Yp & 1.94 $\pm$ 0.18 & \nodata & \nodata \\
MPE13 & G108.57$-$02.76 & 1200 & \nodata & Sh2-151 & Y & 14.90 $\pm$ 0.29 & 39.22 $\pm$ 0.10 & 0.61 $\pm$ 0.01 \\
MPE14 & G108.60+03.15 & 1050 & \nodata & \nodata & Y & 7.78 $\pm$ 0.30 & 24.99 $\pm$ 0.07 & 0.51 $\pm$ 0.02 \\
MPE15 & G108.81$-$01.01 & 270 & \nodata & Sh2-153 & Y & 7.73 $\pm$ 0.07 & 12.71 $\pm$ 0.01 & 0.86 $\pm$ 0.00 \\
MPE16 & G109.07+01.76 & 2800; 1640 & 115 & Sh2-154 & Y & \nodata & \nodata & \nodata \\
MPE17 & G109.14$-$01.58 & 1500 & \nodata & Du 54 & Y & 6.44 $\pm$ 0.40 & 23.81 $\pm$ 0.11 & 0.44 $\pm$ 0.03 \\
MPE18 & G109.32$-$00.50 & 2170 & \nodata & Du 55 & Y & \nodata & \nodata & \nodata \\
MPE19 & G110.46$-$01.43 & 3275; 1100 & 115 & \nodata & Y & \nodata & \nodata & \nodata \\
MPE20 & G110.97+04.31 & 580 & \nodata & \nodata & Y & 2.83 $\pm$ 0.12 & 6.56 $\pm$ 0.03 & 0.68 $\pm$ 0.02 \\
MPE21 & G111.18$-$00.56 & 2900 & \nodata & Sh2-157 & Y & \nodata & \nodata & \nodata \\
MPE22 & G111.25+02.75 & 770 & \nodata & \nodata & Y & 4.37 $\pm$ 0.19 & 22.76 $\pm$ 0.12 & 0.26 $\pm$ 0.02 \\
MPE23 & G112.03+01.16 & 2480 & \nodata & Sh2-161 & Y & \nodata & \nodata & \nodata \\
MPE24 & G112.19+03.79 & 3410; 1630 & 110 & Sh2-160 & Y & \nodata & \nodata & \nodata \\
MPE25 & G113.05+02.16 & 3450; 2200 & 0 & Du 56; Du 57 & Y & \nodata & \nodata & \nodata \\
MPE26 & G113.20$-$00.19 & 830; 450 & 103 & \nodata & Y & \nodata & \nodata & \nodata \\
MPE27 & G113.28+00.53 & 680 & \nodata & \nodata & Y & 2.44 $\pm$ 0.16 & 4.21 $\pm$ 0.05 & 0.83 $\pm$ 0.03 \\
MPE28 & G113.62$-$00.38 & 780 & \nodata & \nodata & Y & 9.03 $\pm$ 0.23 & 26.01 $\pm$ 0.05 & 0.57 $\pm$ 0.01 \\
MPE29 & G114.02$-$02.50 & 3950; 2050 & 110 & Du 58; Du 59 & Y & \nodata & \nodata & \nodata \\
\enddata
\tablenotetext{a}{``Y'': visually detected; ``Yp'': visually detected but partially observed.}
\tablecomments{MIPAPS names are assigned according to the central positions of the individual Pa$\alpha$ features. Radius values approximate the angular extents of the Pa$\alpha$ features, and they were used as aperture radii for the photometries of the Pa$\alpha$ and H$\alpha$ total fluxes. The source with elliptical extent has two radius values, which are the semi-major and semi-minor axes of the ellipse with the given position angle. The yellow dotted circles and ellipses with these central positions, radii, and position angles are denoted in Figures \ref{fig:paa} and \ref{fig:ha}. Corresponding known \ion{H}{2} regions are found in the SIMBAD database.
}
\end{deluxetable*}

\clearpage
\begin{deluxetable*}{cclcr}
\tablecaption{MIPAPS Pa$\alpha$ Point-like Sources\label{table:mpp}}
\tablewidth{700pt}
\tabletypesize{\footnotesize}
\tablehead{
\colhead{ID} & \colhead{MIPAPS Name} & \colhead{Corresponding Known Object (Type)} &
\colhead{IPHAS H$\alpha$ Detection\tablenotemark{a}} & \colhead{Pa$\alpha$ Total Flux} \\
\colhead{} & \colhead{} & \colhead{} & \colhead{} & \colhead{(10$^{-14}$ W m$^{-2}$)}
}
\startdata
MPP01 & G097.98$-$01.03 & MWC 645 (Emission-line Star) & Y & 2.08 $\pm$ 0.05 \\
MPP02 & G098.26+04.91 & PN K 3-60 (Planetary Nebula) & Y & 0.17 $\pm$ 0.03 \\
MPP03 & G099.21$-$01.18 & AS 481 (Emission-line Star) & Y & 0.34 $\pm$ 0.04 \\
MPP04 & G099.53+04.40 & HD 239712 (Emission-line Star) & Y & 0.44 $\pm$ 0.04 \\
MPP05 & G102.66+01.39 & WR 151 (Wolf-Rayet Star) & Y & 0.30 $\pm$ 0.05 \\
MPP06 & G102.78$-$00.65 & WR 153 (Wolf-Rayet Star) & Y & \nodata \\
MPP07 & G103.85$-$01.18 & WR 154 (Wolf-Rayet Star) & Y & \nodata \\
MPP08 & G104.11+01.00 & PN Bl 2-1 (Planetary Nebula) & Y & 0.31 $\pm$ 0.05 \\
MPP09 & G105.32$-$01.29 & WR 155 (Wolf-Rayet Star) & Y & 1.72 $\pm$ 0.04 \\
MPP10 & G106.39+03.09 & V669 Cep (Emission-line Star) & Y(P+D) & 0.40 $\pm$ 0.04 \\
MPP11 & G107.34+04.28 & AS 492 (Emission-line Star) & no data & 0.82 $\pm$ 0.04 \\
MPP12 & G107.51+00.09 & EM* GGR 102 (Emission-line Star) & Y(P+D) & 0.45 $\pm$ 0.04 \\
MPP13 & G107.67+01.40 & MWC 657 (Emission-line Star) & Y & 1.40 $\pm$ 0.04 \\
MPP14 & G107.84+02.32 & NGC 7354 (Planetary Nebula) & Y & 3.01 $\pm$ 0.04 \\
MPP15 & G109.82+00.92 & WR 156 (Wolf-Rayet Star) & no data & 1.86 $\pm$ 0.03 \\
MPP16 & G110.90+01.90 & AS 505 (Herbig Ae/Be Star) & Y & 0.90 $\pm$ 0.05 \\
MPP17 & G111.33$-$00.24 & WR 157 (Wolf-Rayet Star) & Y & 0.83 $\pm$ 0.26 \\
MPP18 & G111.73+00.04 & MWC 1080 (Herbig Ae/Be Star) & Y(P+D) & 2.60 $\pm$ 0.05 \\
\enddata
\tablenotetext{a}{``Y'': visually detected; ``Y(P+D)'': point source and diffuse feature are detected together; ``no data'': not observed.}
\tablecomments{MIPAPS names are assigned according to the central positions of the individual Pa$\alpha$ features. Corresponding known objects are found in the SIMBAD database. All of the Pa$\alpha$ total fluxes were estimated by performing aperture photometry with the same aperture size of 6 arcmin ($\sim$7 MIRIS pixels).
}
\end{deluxetable*}

\clearpage
\begin{deluxetable*}{cccccccl}
\tablecaption{Distances and Information in \citet{foster15} for 33 MIPAPS Pa$\alpha$ Sources\label{table:dis}}
\tablewidth{700pt}
\tabletypesize{\footnotesize}
\tablehead{
\colhead{ID} & \multicolumn{3}{c}{Known Distance} && \multicolumn{3}{c}{Information for Corresponding Sources in \citet{foster15}} \\
\cline{2-4} \cline{6-8}
\colhead{} & \colhead{Kinematic} & \colhead{Spectrophotometric} & \colhead{Parallax} &&
\colhead{Name} & \colhead{$E(\bv)$} & \colhead{Spectral Type of Ionizing Star} \\
\colhead{} & \colhead{(kpc)} & \colhead{(kpc)} & \colhead{(kpc)} && \colhead{} & \colhead{(mag)} & \colhead{}
}
\startdata
WK01 & 9.5 $\pm$ 1.3 & (8.06 $\pm$ 1.61) & \nodata && Sh2-128 & 1.768 & O7V \\
WK03 & \nodata & 1.00 $\pm$ 0.08 & \nodata && Sh2-131 & 0.500 & B1.5V; O6.5V; B1V; B0V; B1V; B2IV; B0.5V; O9.5IV; B0V; B2.5V; B2V; B0.5V \\
WK04 & (5.7 $\pm$ 1.4) & 3.44 $\pm$ 0.29 & \nodata && Sh2-132 & 0.825 & B0V; B0III; O8.5V; O8(V); B0(V); B2(V) \\
WK05 & \nodata & 1.40 $\pm$ 0.28 & \nodata && Sh2-135 & 0.955 & O9.5V \\
WK06 & (5.2 $\pm$ 1.4) & 3.22 $\pm$ 0.29 & \nodata && Sh2-139 & 0.610 & B2III; O8V; B2V \\
WK08 & 7.1 $\pm$ 1.2 & (9.92 $\pm$ 1.98) & \nodata && Sh2-141 & 1.259 & O8V \\
WK09 & (4.1 $\pm$ 1.4) & 3.48 $\pm$ 0.24 & \nodata && Sh2-142 & 0.628 & B0III; B0.5V; B0.2III; B0.5V; O5V; B1V; B1.5V; O8V; B1V \\
WK10 & (4.1 $\pm$ 1.4) & 3.84 $\pm$ 0.77 & \nodata && Sh2-143 & 0.655 & O9.5V \\
WK11 & 6.1 $\pm$ 1.2 & \nodata & \nodata && \nodata & \nodata & \nodata \\
WK17 & (5.3 $\pm$ 1.3) & 2.90 $\pm$ 0.58 & \nodata && Sh2-152 & 1.291 & O8.5V \\
WK19 & (5.6 $\pm$ 1.3) & 2.68 $\pm$ 0.54 & \nodata && Sh2-156 & 1.278 & O6.5V \\
WK20 & \nodata & 0.84 $\pm$ 0.04 & \nodata && Sh2-155 & 0.825 & O9V; O7V; B1V; B0IV; B1V; B1.5V; B3V; B1V; O8.5V; B1V \\
WK21 & 5.1 $\pm$ 1.3 & \nodata & \nodata && Sh2-157 & 0.915\tablenotemark{a} & O9.5V\tablenotemark{a} \\
WK22 & (5.6\tablenotemark{b}) & (2.44 $\pm$ 0.77) & 2.7 $\pm$ 0.1 && Sh2-158 & 1.614 & O9V; O3V \\
WK25 & (4.4 $\pm$ 1.3) & 2.41 $\pm$ 0.16 & \nodata && Sh2-162 & 0.510 & O6.5III; B2III \\
WK26 & (4.4 $\pm$ 1.3) & 3.01 $\pm$ 0.41 & \nodata && Sh2-163 & 1.132 & O9.5V; B1III; O9V; O8V; B1V; B0(V) \\
WK27 & (4.9 $\pm$ 1.3) & 3.08 $\pm$ 0.62 & \nodata && Sh2-164 & 1.020 & B0.2III \\
WK28 & (3.4 $\pm$ 1.3) & 1.96 $\pm$ 0.39 & \nodata && Sh2-165 & 0.760 & B0V \\
WK29 & (4.9 $\pm$ 1.3) & 2.36 $\pm$ 0.47 & \nodata && Sh2-166 & 0.988 & O9.5(V) \\
WK30 & 4.2 $\pm$ 1.3 & \nodata & \nodata && Sh2-168 & 1.025\tablenotemark{c} & O9.5V\tablenotemark{c} \\
WK31 & (4.5 $\pm$ 1.3) & 2.09 $\pm$ 0.42 & \nodata && Sh2-169 & 0.840 & B0.5III \\
WC03 & 9.2 $\pm$ 1.3 & \nodata & \nodata && \nodata & \nodata & \nodata \\
WC10 & 7.4 $\pm$ 1.1 & \nodata & \nodata && \nodata & \nodata & \nodata \\
WC14 & \nodata & 4.87 $\pm$ 0.04 & \nodata && DA 568 & 1.084 & O5V; B2V \\
WC15 & \nodata & 6.18 $\pm$ 1.24 & \nodata && BFS 10 & 1.730 & O9V \\
WC27 & \nodata & 1.63 $\pm$ 0.33 & \nodata && Sh2-134 & 1.038 & B1(V) \\
WC49 & 5.0 $\pm$ 1.4 & \nodata & \nodata && \nodata & \nodata & \nodata \\
WC51 & \nodata & \nodata & 0.7 $\pm$ 0.1 && \nodata & \nodata & \nodata \\
WC54 & 5.8 $\pm$ 1.2 & \nodata & \nodata && \nodata & \nodata & \nodata \\
WG14 & \nodata & 2.96 $\pm$ 0.59 & \nodata && Sh2-161 & 1.200 & B0.5Ib \\
WR72 & 5.8 $\pm$ 1.2 & \nodata & \nodata && \nodata & \nodata & \nodata \\
MPE13 & \nodata & 3.26 $\pm$ 0.26 & \nodata && Sh2-151 & 0.710 & O9.5V; O9.5III; B1.5IV; O9.5V; B2V; B2V; B0.5V; B2V; B2V \\
MPE15 & \nodata & 4.60 $\pm$ 0.41 & \nodata && Sh2-153 & 0.748 & O9.5V; B0.5V \\
\enddata
\tablenotetext{a}{The values are for ALS 19704 in Table 1 of \citet{foster15}.}
\tablenotetext{b}{The value is from \citet{moscadelli09}.}
\tablenotetext{c}{The values are for LS I +6050 in Table 1 of \citet{foster15}.}
\tablecomments{Kinematic distances come from Table 6 of \citet{anderson14}, except one commented on in the footnote. Spectrophotometric distances come from Table 2 of \citet{foster15}. Parallax distances come from \citet{moscadelli09}. For the sources with more than one distance value, distance values that have the highest distance-to-uncertainty ratios are listed without parentheses, and they were applied to Figure \ref{fig:cecorr}(d). For the corresponding sources in \citet{foster15}, names and $E(\bv)$ values (except two commented on in the footnotes) come from their Table 2, and spectral types of ionizing stars come from their Table 1.
}
\end{deluxetable*}
\clearpage
\end{landscape}

\end{document}